\documentclass[10pt]{article}
\usepackage{amsmath,amssymb,graphicx}
\batchmode

\newtheorem{lem}{Lemma}
\newtheorem{thm}{Theorem}

\newtheorem{cor}{Corollary}

\newcommand{\pr}{\noindent{\bf Proof}. }
\newcommand{\re}{\noindent{\bf Remark}. }
\newcommand{\res}{\noindent{\bf Remarks}. }

\newcommand{\pa}{\partial}

\newcommand{\const}{\textrm{const}}

\newcommand{\supp}{ \mathrm{ supp  }}

\newcommand{\al}{\alpha}
\newcommand{\De}{\Delta}
\newcommand{\de}{\delta}
\newcommand{\ga}{\gamma}
\newcommand{\Ga}{\Gamma}
\newcommand{\ka}{\kappa}
\newcommand{\La}{\Lambda}
\newcommand{\la}{\lambda}
\newcommand{\Om}{\Omega}
\newcommand{\om}{\omega}

\newcommand{\ep}{\epsilon}

\newcommand{\Up}{\Upsilon}

\newcommand{\cA}{{\cal A}} 

\newcommand{\cC}{{\cal C}}

\newcommand{\cO}{{\cal O}}

\newcommand{\cH}{{\cal H}}
\newcommand{\cS}{{\cal S}}
\newcommand{\cE}{{\cal E}}

\newcommand{\cT}{{\cal T}}
\newcommand{\cF}{{\cal F}}

\newcommand{\cN}{{\cal N}}

\newcommand{\cJ}{{\cal J}}

\newcommand{\bbR}{{\mathbb{R}}}
\newcommand{\bbZ}{{\mathbb{Z}}}
\newcommand{\bbC}{{\mathbb{C}}}

\begin{document}

\newpage
\title{Transition amplitudes and sewing properties for bosons  on the  Riemann sphere}
\author{ 
J. Dimock\\
Dept. of Mathematics \\
SUNY at Buffalo \\
Buffalo, NY 14260 }
\maketitle

\begin{abstract}
We consider scalar quantum fields on the sphere,
both massive and massless.   
In the massive case we   show that the   correlation functions define amplitudes which are 
trace class operators between tensor products of a fixed  Hilbert space.
We also  establish certain sewing properties between these operators.
In the massless case we consider exponential fields and have a  
conformal field theory.   In this case the amplitudes are only bilinear forms but 
still we establish  sewing properties.  Our results are obtained in a functional
integral framework.  
\end{abstract}

\newpage

\section{Introduction}

A conformal field theory  is specified by  a family of correlation functions defined on a Riemann surface.
These correlation 
functions  can be interpreted  as transition amplitudes   between  various Hilbert spaces 
all built up as  tensor products  of a fixed Hilbert space.    If the conformal field theory is 
describing a statistical mechanics model at the critical point then these amplitudes can be 
thought of as a generalization of the transfer matrix.  If the conformal field theory is 
describing a string theory then these    amplitudes are connected with   scattering amplitudes.
The expected mathematical structure of these amplitudes  was  developed  in a series of
axioms due to G. Segal  \cite{Seg89}, \cite{Seg04},   \cite{Gaw99}.   Verification of
the axioms has been slow with  the best results  obtained   for fermions   \cite{Seg04}, \cite{Hua97}.
  In the present paper we  make some progress   on    verification for bosons in the case where the Riemann surface is a sphere.

We work  in a functional integral formulation.
Because  conformal field theories are massless  the functional integrals are somewhat 
singular and the manipulations one would like to make  are awkward.   Things 
are better for massive fields and so we start with this case.  Then the fields 
satisfy a Markov property  \cite{Nel73c},  \cite{Dim04} which makes it possible to reduce
certain   integrals over fields on the whole sphere to integrals over fields   on one-dimensional submanifolds.  This   property facilitates the definition of the amplitudes and the sewing properties.     We develop this massive case at length.

In the massless case one does not have a  Markov property,  at least not in the same  strong sense as
  in   the massive case.
The original idea was to carry over    results from   the massive case by  taking the limit as the 
mass goes to zero.  Unfortunately  many of the massive results  do not hold for small mass,  let alone
uniformly in the mass.  So  for the moment at least  this  strategy    is not as  rewarding as    one might have hoped.  

 One  property that  does   carry over to the massless case    is a   reflection positivity result.    Taking advantage of this  and using  some simplifications  due to the conformal symmetry we are able to define amplitudes   and establish sewing properties in this  case  as well.  The results  are  somewhat weaker  than in the massive case.   The results for both cases are described in more detail in section \ref{bunko}.

\section{Preliminaries}

\subsection{metrics}  \label{metrics}

We work on the Riemann sphere  $\bbC_{\infty}  =  \bbC  \cup   \{ \infty \}$.     Complex  coordinates are
the standard  $z=x_1+ix_2$ on  $\bbC$ and  
  $\zeta$  on  $\bbC_{\infty} - \{ 0\}$  which is   $\zeta  =  1/z$ on the overlap.  We also label points  in $\bbC$
  by  $x  =  (x_1,x_2)$ when we want to ignore the complex structure.  We will need to refer to
 the unit  circle  $C_0$ and the closed  regions   that it bounds:
\begin{equation}
\begin{split}
C_0  =&  \{  z \in \bbC_{\infty}:   |z| =1\} \\ 
D_0  =&  \{  z \in \bbC_{\infty}:   |z|  \leq 1\} \\ 
D'_0  =&  \{  z \in \bbC_{\infty}:   |z|  \geq  1\} \\ 
\end{split}
\end{equation}

We consider   conformal    metrics on  $\bbC_{\infty}$ which have the form   in $\bbC$ 
\begin{equation}
 \ga  =  \rho(z)  |dz|^2 =   \rho(x)  ( dx_1^2 + dx_2^2 )
\end{equation}
for some smooth positive function  $\rho$.
In the other patch $\zeta =1/z$  it has the form  
$\ga  =   |\zeta|^{-4} \rho(1/\zeta)
 | d \zeta|^2$   so
 $  |\zeta|^{-4} \rho(1/\zeta) $   should  be smooth and positive at  $\zeta=0$.

For a Hilbert space structure in our field theory we will    want to consider metrics $\ga$   which are invariant under  radial reflection through
$C_0$.    Radial reflection   is defined by   
\begin{equation} 
\theta(z)   =  \bar z^{-1}   = \frac { z }{|z|^2}
\end{equation}
which   preserves  $C_0$  and   exchanges  $D_0$ and  $D'_0$.
We    want $ \theta^* \ga  =  \ga$   and if   $\ga  = \rho|dz|^2$  the condition is
that    
\begin{equation}   \label{boot}
 |z|^{-4}\rho(\bar z^{-1}  )   =    \rho(z)
\end{equation}   
A reflection invariant metric is  the
round  metric 
\begin{equation}
\ga  =   \frac{4}{(1+|z|^2)^2} |dz|^2
\end{equation}
Another  reflection invariant metric  is  the cylindrical metric
\begin{equation}
\ga =  \frac{1}{|z|^2}  |dz|^2
\end{equation}
This is  actually   not  a metric on the whole   sphere  but  only  on  $\bbC- \{ 0\}$.   Under the
mapping     $z=e^{iw}$
this   is identified with  the  flat    metric  $|d w|^2 $ on the cylinder      $(\bbR/2\pi \bbZ)   \times  \bbR $.

As a point of reference  we will  pick a standard metric.  A metric  $\ga_0  =  \rho_0(z)  |dz|^2$   is defined to be a  \textit{standard metric}   if it is     invariant  under  
 radial  reflections  and also   rotations
(i.e.   $\rho_0(e^{i\theta} z)  =  \rho_0(z)$)  and if  there is a constant  $d$ such that   
 that   it  has the   toroidal  form   $|z|^{-2}  |dz|^2$   for   $e^{-d}  <  |z|  <  e^{d}$.   The last
 requirement is to keep things simple in a neighborhood of  $C_0$.    Note that 
 with this metric the strip   $e^{-d}  <  |z|  <  e^{d}$
has width $2d$.

\subsection{Laplacians}
    Associated with any conformal  metric  $ \ga  =  \rho(z)  |dz|^2 $  on  $\bbC_{\infty}$ we have  a measure ($|\ga| = \det \ga$)
\begin{equation}
d \mu_{\ga}(x)  = | \ga(x)|^{1/2} dx   =      \rho(x)  dx
\end{equation}
and the Hilbert space   $L^2(\bbC_{\infty},   \mu_{\ga})$
with the inner product   $(f,h)_{\ga}  =  \int  \bar f h d \mu_{\ga}$
The   Laplacian  for this metric is   
\begin{equation}
\De_{\ga}   =  \frac{4}{\rho(z)}  \frac{ \pa^2}{\pa z \pa  \bar z}  =  \frac{1}{\rho(x)}  \left(  \frac{\pa^2}{\pa x_1^2}  +  \frac{\pa^2}{\pa x_2^2}\right)
\end{equation}
The negative  Laplacian  $-\De_{\ga}$   is naturally  a  positive self-adjoint operator  on
 $L^2(\bbC_{\infty},  d \mu_{\ga})$
and has purely discrete spectrum.    The lowest eigenvalue is $0$  and the eigenfunctions 
are the constants.
 For     $\mu  >0$   the operator 
$( - \De _{\ga}  + \mu)^{-1}$ exists and    is  
 Hilbert-Schmidt. 
 This  is true    on any  compact two-dimensional  manifold,  see for example 
\cite{Shu01}, p. 113.

 We also need  Sobolev spaces.  For any real number  $s$  the Sobolev space
 $H^s$ is a space of distributions  on $\bbC_{\infty}$  defined by the  requirement that
 in local  coordinates  they be in the corresponding  Sobolev space  on   $\bbR^2$. 
  We   can  give    $H^s$   a  norm  and regard it   as a Hilbert space  by an  alternate definition.
 For any  $\ga, \mu$    define  $H^{s}_{\ga, \mu}$  to be the completion  of 
  $\cC^{\infty}(\bbC_{\infty})$
  in the norm \begin{equation}
  \|f\|_{s, \ga, \mu}  =   (f,  ( - \De _{\ga}  + \mu)^{s} f)_{\ga}^{1/2}
  \end{equation} 
  Then  $H^s= H^s_{\ga,\mu}$  as a vector spaces.
  We have   $H^{0}_{\ga,\mu} =  L^2( \bbC_{\infty},  d  \mu_{\ga})$  and we have the 
  inclusions
  $  H^{+1}_{\ga,\mu}    \subset   H^0_{\ga,\mu}    \subset   H^{-1}_{\ga,\mu}   $.  
  The inner product   $(f,h)_{\ga} $ extends to a bilinear form  on  $\cH^{+1}  \times  \cH^{-1}$
  and to emphasize this interpretation we sometimes write it as    $(f,h)_{+1,-1}$.

\newpage

\section{Massive fields}

\subsection{fields}  \label{singing}

Now we define massive scalar fields  on  the sphere  $(\bbC_{\infty},  \ga )$  with an arbitrary
metric $\ga$.   As the test   function  space  we  take
the real  Sobolev space   $ H^{-1}_{\ga,\mu }$. 
Let     $\{ \phi(f) \} $  with  $f \in  H^{-1}_{\ga,\mu}$ be a family of Gaussian random variables with
covariance  given by the inner product.  These  are functions on an underlying probability  space   $(Q, \Sigma, m_{\ga,\mu})$.
 Expectations are denoted by  $< \cdots>_{\ga,\mu} $ so we have the characteristic function
\begin{equation}
<  e^{i \phi(f)}>_{\ga,\mu}   = \int_Q  e^{i \phi(f)} dm_{\ga,\mu}  = 
   \exp  \left(  -  \frac12   \|f\|^2_{-1, \ga, \mu}   \right)
\end{equation}
The family  $\{ \phi(f) \} $ is  our quantum field theory with mass   $\sqrt{\mu} >  0$.

We introduce Wick-ordered products in the standard way defining   $:\phi(f_1)  \cdots  \phi(f_n):$
to be the projection in $L^2(Q, \Sigma, m_{\ga,\mu})$
     of  $\phi(f_1)  \cdots  \phi(f_n)$  onto   the orthogonal complement of 
polynomials   in $\phi(f)$    of degree $n-1$.   Then   $:\phi(f_1)  \cdots  \phi(f_n):$ is a polynomial
of degree $n$   and all such polynomials   span a dense set  in  $L^2$.     Any  contraction
$T$ on    $H^{-1}_{\ga,\mu }$  induces a contraction  $\Ga(T)$ on 
$ L^2(Q,   \Sigma ,  m_{\ga,m})$   which satisfies  $\Ga(T)1 =1$ and
\begin{equation}  \label{second}
\Ga(T)   :\phi(f_1)  \cdots  \phi(f_n):
=  :\phi(Tf_1)  \cdots  \phi(Tf_n):
\end{equation}
  We have   $\Ga(T) \Ga(S)  = \Ga(TS)$  and  $\Ga(A)^*  =  \Ga(A^*)$.  If  $U$ is
  unitary then 
  $\Ga(U)   \phi(f ) \Ga(U^{-1})=  \phi(Uf) $,
but not in general.

For   any  closed subset  $A  \subset  \bbC_{\infty}$  let 
   $\Sigma_A$   be the    $\sigma$-algebra   of measurable subsets generated by   the random variables
    $\{ \phi(f)  \}$       with  $\supp  f   \subset  A$.   Let  $\cE^{\ga, \mu}_A$ denote the   conditional expectation  with respect to   $\Sigma_A$ for  the measure  $m_{\ga, \mu}$.
As an operator on   $L^2(Q, \Sigma, m_{\ga,\mu})$,
$\cE^{\ga, \mu}_A$ can be characterized as  the projection    $\cE^{\ga, \mu}_A  =  \Ga(e^{\ga, \mu}_A)$  where    $e^{\ga, \mu}_A$   is the projection
 in    $H^{-1}_{\ga, \mu}$  onto    elements with support in  $A$.  \cite{Si75}

 The fields have the   the \textit{Markov property}   \cite{Nel73b}, \cite{Dim04} which states
  that for an open set  $\Om \subset  \bbC_{\infty}$
 \begin{equation}
 \cE^{\ga, \mu}_{\Om^c}\cE^{\ga, \mu}_{\overline{ \Om}}   =   \cE^{\ga, \mu}_{\pa \Om}
\end{equation} 
Another way to put it is  that if  $F$ is measurable with respect to  $\Sigma_{\overline{\Om}}$
then    $ \cE^{\ga, \mu}_{\Om^c}F   =   \cE^{\ga, \mu}_{\pa \Om}F  $.

 \subsection{a standard  Hilbert  space}      \label{nuts}
 
 We   now pick  a  fixed  standard  metric  $\ga_0$ as defined in section \ref{metrics}.   The invariance
 under radial reflections gives a reflection positivity property  for the fields.    The  latter is
 used to create a  standard  Hilbert space.   
 Reflection positivity has long played a key role in   field theory on Riemannian manifolds
 \cite{OS73},     \cite{Nel73c},    \cite{DDD86},  \cite{Dim04},  \cite{JaRi06}.

     Because radial   reflection  $\theta$   is an isometry it induces  
an unitary operator  $\theta^*$   on   $H^{-1}_{\ga_0,\mu}$   and hence a unitary
operator   $ \Ga(\theta^*)$ on   $ L^2(Q,   \Sigma ,   m_{\ga_0,\mu})$.
We  define      
\begin{equation}  \label{green}
\Theta  \Psi   =  \overline {  \Ga(\theta^*) \Psi  }
\end{equation}
Then  $\Theta$   is anti-unitary  and satisfies   $\Theta^2  =1$.  
Now  let     $\Psi  \in    L^2(Q,   \Sigma ,   d m_{\gamma_0,\mu})$ be
 measurable with respect to   $\Sigma_{  D_0}$,  written 
   $\Psi \in   L^2(Q,   \Sigma_{  D_0} ,    m_{\ga_0,\mu})$. 
    Then  $\Theta \Psi$ is measurable 
with respect to   $\Sigma_{ D^c_0}$ and using the Markov  property   we find 
the  reflection   positivity result    \cite{Nel73c}, \cite{Dim04}
\begin{equation}  \label{sing}
<(\Theta \Psi) \Psi>_{\ga_0,\mu}  =   
    \int   |  \cE_{C_0}  \Psi  |^2     d m_{\gamma_0,\mu }  \geq  0
\end{equation}
Using this we   give three equivalent  constructions  of  a standard Hilbert space  $\cH$.
\bigskip

\begin{enumerate}
\item   \textbf{first construction}.   
The first construction  is  analagous   to   the Osterwalder-Schrader  reconstruction
theorem  for Euclidean field theory.    Consider the multilinear functional 
from     $H^{-1}  \times  \cdots  \times  H^{-1}$ to  random variables on  
$(Q, \Sigma,  m_{\ga_0, \mu})$   which 
sends   $(f_1,  \dots, f_n)  \to    \phi(f_1)  \cdots  \phi(f_n)$.
This induces  a map    $\Phi_n$  from   the algebraic tensor 
product
$ \otimes_{i=1}^n   H^{-1}$  to   random variables such that
\begin{equation}
\Phi_n(f_1  \otimes   \cdots   \otimes   f_n)
=   \phi(f_1)  \cdots  \phi(f_n)
\end{equation}
This is the universal property of the algebraic tensor product   \cite{Gre78}. 
Next  consider  the  vector  space of all  finite  sequences
\begin{equation}
F= (F_0, F_1, F_2,  \dots  )  \hspace{1cm}    F_n   \in   \otimes_{i=1}^n   H^{-1}
\end{equation}
and let   $\cS$ be the complexification.
Define a linear map from  $\cS$ to random variables by   
\begin{equation}
\Phi(F)   =   \sum_n   \Phi_n(F_n)
\end{equation}
where  $\Phi(F_0)  =  F_0  \in \bbC$.  The map  $\theta^*$  on  $H^{-1}$ induces a 
map  on   sequences   $F$   and we  let   $\Theta$  on  $\cS$  be this map followed by complex conjugation.
The  new  map  $ F  \to \Theta F $  is related to  the previous definition
 in    (\ref{green})  by   $\Phi(\Theta F)  =  \Theta \Phi(F)$.

Next   for    any  closed   subset   $A \subset  \bbC_{\infty}$  we let   $\cS_A$
be sequences in which all functions have support in  $A$.
   We  want to define   a  norm     on  $\cS_{D_0}$ by 
\begin{equation}
\|  F \|^2 
 = <   \Phi(\Theta F) \Phi(F)  >_{\ga_0, \mu}  
 =    \sum_{n,m}     <   \Phi_n (\Theta F_n)  \Phi_m(F_m)>_{\ga_0, \mu }
\end{equation}
Since     $  <   \Phi(\Theta F)  \Phi(F) >_{\ga_0, \mu}  = < \Theta ( \Phi(F))  \Phi(F) >_{\ga_0, \mu} $
this  is non-negative by  (\ref{sing}).    But it is not definite.
 We   divide   by the null space   $\cN  = \{   F  \in  \cS_{D_0} :   \|F\| =0\}      $  to get  a pre-Hilbert space   
$  \cH_0  =   \cS  _{D_0}/  \cN$.  Then   complete  it   to get  a Hilbert space (depending on 
$\ga_0, \mu$)   
\begin{equation}  \label{scotch}
\cH   =   \overline{  \cH_0} =  \overline{ \cS_{D_0}/   \cN }
\end{equation}
Let     $\nu $ map   an element of    $  \cS  _{D_0}$  to  its equivalence class in  $ \cH_0  \subset \cH$.
Then 
\begin{equation}
  (\nu(F_1), \nu(F_2))    =   <    \Phi (\Theta F_1)\Phi( F_2)>_{\ga_0,\mu}  
\end{equation}

\item   \textbf{second construction}.   
A second construction is a variation of this in which the $L^2$  space plays a more prominent role.  Consider the Hilbert space  $ L^2(Q,   \Sigma_{D_0} ,   m_{\gamma_0,\mu})$   but 
now supplied with the norm    $\|\Psi \|^2   =  <(\Theta \Psi)  \Psi>_{\ga,\mu}$
and  denoted    $ L^2_{\Theta}(Q,   \Sigma_{D_0} ,    m_{\gamma_0,\mu})$   .
Divide   
by   the null  space    $\cN=   \{  \Psi: \|\Psi \|  =0 \}  $   and get   $\cH'_0=L^2_{\Theta}(Q,   \Sigma_{D_0} ,    m_{\gamma_0,\mu})/  \cN  $.   Then complete it   to obtain 
the    Hilbert space
\begin{equation}  \label{sloppy}
\cH   = \overline {\cH'_0} =   \overline{  L^2_{\Theta}(Q,   \Sigma_{D_0} ,    m_{\gamma_0,\mu})/  \cN  }
\end{equation}
Let      $\nu $   map an element of    $ L^2_{\Theta}(Q,   \Sigma_{D_0} ,    m_{\gamma_0,\mu})$ to
its equivalence class in  $\cH'_0 \subset  \cH$.
Then   
\begin{equation}
( \nu (\Psi_1), \nu(\Psi_2))  =   <(\Theta \Psi_1)  \Psi_2>_{\ga_0,\mu}  
\end{equation}

To see that this construction is equivalent to the first  note that the 
 map   $F  \to   \Phi(F)$    from    $\cS_{D_0}$  to    
 $ L^2_{\Theta}(Q,   \Sigma_{D_0} ,    m_{\gamma_0,\mu})$  is norm preserving
 and so determines a norm preserving map    from   $\cH_0$ to   $\cH'_0$   which takes   $\nu(F)$ to  
 $ \nu(\Phi(F))$.    We argue that the range is dense hence   hence    the map   extends to a     unitary   from     $\cH$   as defined   (\ref{scotch}) in   to  $\cH$
as defined in   (\ref{sloppy}) .

Now   polynomials  $\Phi(F)$,  $F \in \cS$  are dense in      $ L^2(Q,   \Sigma ,    m_{\gamma_0,\mu})$.  ( More  precisely  we can choose the measure space so this is true.) 
  Hence    polynomials  $\Phi(F)$,  $F   \in \cS_{D_0}$  are dense in      $ L^2(Q,   \Sigma_{D_0} ,    m_{\gamma_0,\mu})$  and hence they are dense in 
  $ L^2_{\Theta}(Q,   \Sigma_{D_0} ,    m_{\gamma_0,\mu})$.  
 Hence    vectors  $\nu(\Phi(F))$  are dense  in  $\cH'_0$ as required.

     \item   \textbf{third construction}.   
     The third  construction is just to take
     \begin{equation}
\cH   =  L^2(Q,   \Sigma_{C_0} ,    m_{\gamma_0,\mu}) 
  \end{equation}
as the standard space.  To see that this is equivalent
   note that  
the  identity   (\ref{sing})  shows that the map  $ F     \to   \cE_{C_0}F$   
from    
 $ L^2_{\Theta}(Q,   \Sigma_{D_0} ,    m_{\gamma_0,\mu})$
   to   $ L^2(Q,   \Sigma_{C_0} ,    m_{\gamma_0,\mu})$ is norm preserving.   Hence it  defines an isometry from    $\cH'_0 $  to  
     $ L^2(Q,   \Sigma_{C_0} ,   m_{\gamma_0, \mu  })$ which takes  $\nu(\Psi)$ to    $\cE_{C_0}  \Psi$.   But the map is also onto.
  Hence   $\cH'_0 $   is  complete 
  (i.e.  the completion in  (\ref{sloppy}) was unnecessary).   Thus we   have a unitary operator
 from  $\cH$ defined in  (\ref{sloppy})  to  the new space 
 $   L^2(Q,   \Sigma_{C_0} ,    m_{\gamma_0,\mu}) $.
\end{enumerate}

\subsection{the problem} 
 \label{bunko}
We are concerned with the following situation.
Let  $D_1, \dots, D_n$  be a  collection of  disjoint  discs  in  $\bbC_{\infty}$
For each  $i$ we  pick a    Mobius transformation $\al_i$ which takes
$D_i$  to  the unit disc   $D_0$.  We   have 
\begin{equation}  \label{butter}
\al_i (z) =   \left\{  \begin{array}{rcl}
  (z-a_i)/r_i&  &  \textrm{ if }   D_i =      \{ z:  |z- a_i| \leq r_i\}  \\
   r_i/(z-a_i)&  &  \textrm{ if }   D_i =      \{ z:  |z- a_i| \geq r_i\}  
   \end{array}
   \right.
   \end{equation}
A third possibility, which we have  not written  explicitly, is that  $D_i$ is a half plane.
The  pull back of  our standard metric
  $\ga_0=  \rho_0(z) | dz|^2$   is    metric  $\ga_i =  \al_i^*(\ga_0)$  which in  coordinates  $z_i  = \al_i (z)$ is given by  
\begin{equation}
\ga_{i}   =    \rho_0(z_i) |dz_i|^2
\end{equation}

Now using   a partition of unity  construct  metrics $\ga$  with the property
that   $\ga =  \ga_{i}$   on a neighborhood of  $D_i$  and  is
arbitrary elsewhere.    Specifically  define \begin{equation}
\begin{array}{rcl}
D_0  = \{z:  |z| \leq 1   \} &       C_0 =  \{z:   |z| =1  \}      & D'_0  = \{z:  |z| \geq 1   \}  \\
D_{0 +} = \{z:  |z|   \leq  e^{d/2} \}   &       C_{0+} =  \{z:   |z| =e^{d/2}  \}   &   D'_{0 +} = \{z:  |z|   \geq  e^{d/2} \}  \\
D_{0 ++} = \{z:  |z|   \leq  e^{d} \}   &       C_{0++} =  \{z:   |z| =e^{d}  \}   &   D'_{0 ++} = \{z:  |z|   \geq  e^{d} \}  \\
\end{array}
\end{equation}
and let   $D_i, D_{i+}, D_{i++},  C_i,  C_{i+}, C_{i++},D'_i, D'_{i+}, D'_{i++}$  etc.  be  the image of these sets under   $\al_i^{-1}$.
Then     $D_{i+}, D_{i++}$  are enlargments of  $D_i$.   The condition is  that the   $D_{i++}$   are disjoint and  that    $\ga =  \ga_i$  on   $D_{i++}$.
For such a metric $\ga$  we call $(\bbC_{\infty},  \ga)$
a  sphere with  standard  discs $\{D_i\}$.    

We   also  suppose we have a  specific   parametrization of  
of   $D_i$.   Each disc is   labeled as either  an  \textit{in-disc}  or  an   \textit{out-disc}.  
If  $D_i$ is an in disc then  we define  
 \begin{equation}  \label{soon}
j_i (z)   = e^{i\theta_i}\al_i(z) 
\end{equation}
which maps     $  D_i$  to  $D_0$. 
 If  $D_i$ is an out-disc then we define  
 \begin{equation}
j'_i (z)   = e^{i\theta_i}\al_i(z)^{-1}
\end{equation}
which maps   $D_i$ to  $D'_0$.
 In either case   we have allowed  a a twist with the phase   factor  $e^{i\theta_i}$.  Even with the twist we still have   $j_i^*\ga_0 =  \ga_i$ and    $(j'_i)^*\ga_0 =  \ga_i$   thanks to the rotation invariance .  
When supplied with a choice of  maps  $j_i,j'_i$    a  sphere with  standard    discs $\{D_i\}$
said to be parametrized.

Note that  the   
map   $j_i$  induces a pull back      $(j_i^* f) (z) =  f(j_iz)$  on functions or on  $H^{-1}$.  
This     induces   a map    $\cJ_i $ on  $\cS$.  Similarly    $j'_i$ induces a map  $\cJ'_i$
on  $\cS$.    We   have that  
 \begin{equation}
 \begin{split}
&\cJ_i :     \cS_{D_0}  \to  \cS_{D_i}      \\
&\cJ'_i :   \cS_{D'_0}  \to  \cS_{D_i}  \\
 \end{split}
\end{equation}

Now we can state the problem,  more or less   as  posed by  Gawedski \cite{Gaw99}.  Let  $(\bbC_{\infty}, \ga) $ be a  sphere  with standard discs
parametrized so  that  
 $\{D_i\}_{i \in I}$  are  in-discs    and  $\{D_i\}_{i \in I'}$ are out-discs.
 Let  $\{F_i\}_{i \in  I' \cup  I}$ be a   collections of  elements in   $  \cS_{D_0} $.
Then   $\cJ_iF_i  \in \cS_{D_i}$ for  $i \in I$
and   $\cJ_i' \Theta  F_i  \in  \cS_{D_i}$  for   $i \in  I'$ and we can  consider
   \begin{equation}      \label{column} 
<\prod_{i\in I'}\Phi( \cJ'_i \Theta  F_i)    \prod_{i\in I} \Phi( \cJ_iF_i)  >_{\ga,\mu}
\end{equation}
We would like to   establish the following:
\begin{enumerate}
\item  
The expectation  (\ref{column})  depends only on the equivalence class of   $F_i$ and so defines a multilinear
functional on   $\cH_0  \times  \cdots  \times  \cH_0$.
\item  The expectation extends by continuity to a 
multilinear   functional  on  $\cH \times  \cdots  \times  \cH$
\item   The  multilinear    functional defines an operator  $ A^{I'I}_{\ga,\mu} $
from   $[\otimes_{i \in I} \cH]$  to    $[\otimes_{i \in I'} \cH]$
such that    if  $\cF_i  = \nu(F_i)$, etc.
\begin{equation}
\left( [\otimes_{i \in I'}  \cF_i],   A^{I'I}_{\ga,\mu}  [\otimes_{i \in I} \cF_i ]\right) = Z_{\ga, \mu}
  <\prod_{i\in I'}\Phi( \cJ'_i \Theta  F_i)    \prod_{i \in I} \Phi( \cJ_iF_i)  >_{\ga,\mu}
\end{equation}
with a constant  $Z_{\ga, \mu}$ to be specified.
\item   The constants  $Z_{\ga, \mu}$  can be chosen so  certain sewing properties hold, for example 
\begin{equation}
\cA^{I'1}\cA^{1I}  =  \cA^{I'I}
\end{equation}  
\end{enumerate}

We will be able to establish all this is the massive case.  
In the massless case the formulation of the problem is somewhat different,  but 
still there are questions analagous to  these  four.   In this case we establish   (1.) and  weaker versions of (2.),(3.),(4.).

\subsection{measure theory results}

 We start with some preliminary results.  Consider  conformal metrics
  $\ga'  =  \rho' |dz|^2$  and   $\ga  = \rho|dz|^2$.  Then    $\ga' = \la \ga$
where    $\la  =  \rho'/ \rho$ is a smooth function  on  $\bbC_{\infty}$.   By the compactness the positive functions   $\rho, \rho'$  are bounded above and below  and hence so  is  $\la$.     Any function  $\la$ on the sphere   gives a    map  on functions   $f  \to  \la  f$
and this induces a  map  $F  \to  F_{ \la} $  on  $\cS$.

\begin{lem}  \label{slam}
If   $\ga' =  \la \ga$  then for    any  $F  \in  \cS$
\begin{equation}
   <   \Phi(F) > _{\ga', \mu}  
=    <   \Phi(F_{\la}) > _{\ga, \la \mu}
\end{equation}
\end{lem}
\bigskip

\pr  
  Using    $\De_{ \ga'}  =  \la^{-1} \De_{\ga}$ and 
$d \mu_{ \ga'}  =   \la  d\mu_{\ga}$  we  have  for smooth functions $f$
\begin{equation}
 (f,  ( - \De _{\ga'}  +\mu)^{-1} f)_{\ga'} 
=       (\la f,  ( - \De _{\ga } +  \la  \mu)^{-1}\la f)_{\ga }
\end{equation}
Hence the map  $f \to  \la f$ extends to an unitary  from $H^{-1}_{\ga', \mu}$  to  
$H^{-1}_{\ga,  \la \mu}$.    
Then we  have  
\begin{equation}
  <   e^{i\phi(f)} >_{\ga', \mu}  
=   \exp \left( - \frac12   \| f\|^2_{-1,\ga', \mu}  \right)
=   \exp \left( - \frac12   \|  \la  f\|^2_{-1,\ga,  \la  \mu}  \right)
 = <  e^{i\phi(\la f)}  >_{\ga,  \la \mu}  
\end{equation}
Taking derivatives of   the characteristic functions gives a result 
for  polynomials which is what we want.
\bigskip

For the next result   consider     functions of the form  
\begin{equation}
:\phi^2:(g)    \equiv  \int  :\phi(x)^2:  g(x)\    d  \mu_{\ga}(x)
\end{equation}
for some smooth function $g$.
Here  $\phi(x)  =  \phi(\de_x)$  is defined with the  delta   function  $\de_x$.    Since    $\de_x$  is not in  $H^{-1}$,
 the expression $:\phi^2:(g)   $ is not obviously well-defined.       Nevertheless   it does 
define a function  in  $L^2(Q,, \Sigma,  m_{\ga,\mu})$   (and more generally
in  $L^p$ for   $p < \infty$).   This is a standard result in the plane and
we give a treatment for the sphere in Appendix   \ref{A}.

\begin{lem}   { \   }  \label{bam}  For any  smooth positive function    $\la$
  on $\bbC_{\infty}$   
  \begin{enumerate}
\item    $   \int   \exp  \left(  - \frac12      :\phi^2:(\la\mu-\mu)  \right)  
d m_{\ga, \mu}  $  is finite  and non-zero.
\item  $ m_{\ga, \la\mu}$ 
is absolutely continuous with respect to   $m_{\ga,  \mu}$ with   
 Radon-Nikodym derivative  
\begin{equation}    \label{RN}
\left[\frac{   d m_{\ga, \la\mu}}{ d m_{\ga, \mu}}
\right]  =\frac{  \exp  \left(  - \frac12  :\phi^2:(\la\mu-\mu)          \right)  }
{   \int   \exp  \left(  - \frac12      :\phi^2:(\la\mu-\mu)  \right)  
d m_{\ga, \mu}}
\end{equation}
\item  
Let   $\de     =  \inf_{x } \la(x)$.  
\begin{enumerate}
\item  If   $\de \geq 1$  then  
$[  d  m_{\ga, \la \mu}/ d m_{\ga, \mu}] \in L^q$ for   $1 \leq  q<\infty$. 
\item  If
$\de <1$ then   $[  d  m_{\ga, \la \mu}/ d m_{\ga, \mu}] \in L^q$  for     $1\leq   q <  1/(1-\de)$ 
\end{enumerate}
\item   If  $A$  is a closed set and   $\la =1$  on   $A^c$  then  
$[  d  m_{\ga, \la \mu}/ d m_{\ga, \mu}]$   is  $\Sigma_A$
measurable.
  \end{enumerate}
\end{lem}
\bigskip

\re   We  have not insisted  that   $m_{\ga, \la \mu}$  and  $m_{\ga,  \mu}$
are defined on the same space so the    second statement needs some explanation.
The claim is that  if  $\{\phi(f)\}$  is  a Gaussian family  with covariance   
$( - \De_{\ga}   +   \mu) ^{-1}$    on   a measure space 
$(Q,  \Sigma,  m_{\ga,\mu})$   then changing     the measure to  
 $[d  m_{\ga, \la \mu}/ d m_{\ga, \mu}]     d m_{\ga, \mu} $   as   specified by (\ref{RN})  makes  $\{\phi(f)\}$ into   a Gaussian family 
 with covariance   
$( - \De_{\ga}   + \la  \mu) ^{-1}$.      
 \bigskip

\pr    
As explained in   appendix \ref{B}  the first  point is true  if 
\begin{equation}
( - \De_{\ga}   +   \mu)   +    (\la  \mu-\mu)   =  - \De_{\ga}   +  \la \mu   >0
\end{equation}
which is clear since  $\la$ is positive.  Also from  Appendix  \ref{B}      the new measure 
 $[d  m_{\ga, \la \mu}/ d m_{\ga, \mu}]     d m_{\ga, \mu} $    has the claimed  covariance  
 \begin{equation}
  (( - \De _{\ga} +  \mu ) +   (\la \mu -\mu))^{-1}  =  ( - \De _{\ga} + \la \mu)^{-1}
\end{equation}
 The  third      point   is also an integrability  question.    Now we need     
\begin{equation}
( - \De_{\ga}   +   \mu)  + q (\la \mu-\mu) >0
\end{equation}
This is clear if  $\de  \geq  1$  and if  $\de <1$
it   follows from    $\la >   1- 1/q$  which is implied by our condition $  q <  1/(1-\de)$.
The fourth  point follows from the fact that if   $\supp \ g  \subset A$  then   $: \phi^2:(g)$  is   
$\Sigma_A$  measurable; see  lemma \ref{omega} in Appendix \ref{A}. This completes
the proof.
\bigskip

\begin{lem}  \label{thanks}  
 Suppose that    $\ga'  = \la \ga$
 with   $\la =1$  on  $\Om \subset \bbC_{\infty}$  open.
  Then for     $F \in  \cS_{ \bar  \Om}$  
\begin{equation}    \cE_{\pa  \Om}^{\ga', \mu} \Phi( F) =
 \cE_{\pa \Om  }^{ \ga,\la \mu} \Phi( F)
=  \cE_{\pa \Om}^{ \ga,\mu} \Phi(F) 
\end{equation}
\end{lem}
\bigskip

\re  Thus the conditional expectation does not depend on the metric or the mass outside  $\Om$.

Note that  
$\Phi(F)$  is  $\Sigma_{\bar \Om}$  measurable and so by the  Markov property 
\footnote{ The Markov property  is also true with variable mass}  
an equivalent statement is  
\begin{equation}     \label{suzy}
    \cE_{  \Om^c}^{\ga', \mu} \Phi( F) =
 \cE_{ \Om^c  }^{ \ga,\la \mu} \Phi( F)
=  \cE_{ \Om^c}^{ \ga,\mu} \Phi(F) 
\end{equation}
\bigskip

\pr We prove  (\ref{suzy}).
   For any     $G  \in    \cS_{\Om^c}$  we have  that   $\Phi(G)$  is   $\Sigma_{\Om^c}$
measurable and so by  lemma \ref{slam}  we compute 
\begin{equation}
\begin{split} 
  \int   \Phi(G)  \left(  \cE_{\Om^c}^{ \ga', \mu} \Phi(F)  \right) d m_{\ga', \mu} 
=&    \int    \Phi(G)   \Phi(F)   d m_{ \ga',   \mu}       \\
=&    \int    \Phi(G_{\la})   \Phi(F)  dm_{  \ga,  \la \mu}   \\
=&    \int    \Phi(G_{\la})   (  \cE_{\Om^c}^{ \ga,\la \mu}  \Phi(F)  )  d m_{ \ga,  \la \mu}      \\
=&    \int    \Phi(G)   (  \cE_{\Om^c}^{ \ga, \la  \mu} \Phi(F) )  d m_{\ga',  \mu}  \\
\end{split}
\end{equation} 
In the last step we use   again that  $  \cE_{\Om^c}^{ \ga,\la \mu}  \Phi(F) 
=  \cE_{\pa \Om}^{ \ga,\la \mu}  \Phi(F) $  to conclude that the term is unaffected.
Since  polynomials     $\Phi(G)$  are dense in    $L^2(Q , \Sigma_{\Om^c},    m_{ \ga',  \mu} )$
the first identity follows. 

For the second point we  again take   $G  \in    \cS_{\Om^c}$   and 
compute by  lemma \ref{bam}
\begin{equation}
\begin{split} 
\int \Phi(G)   \left(  \cE_{\Om^c}^{ \ga,\la \mu} \Phi(F)  \right) d m_{  \ga, \la  \mu} 
=&    \int  \Phi(G )  \Phi(F)   dm_{ \ga, \la \mu}       \\
=&    \int  \Phi( G ) \Phi(F)  \left[\frac{   d m_{  \ga, \la \mu} }{ d m_{  \ga,  \mu} }
\right]   d m_{  \ga, \mu} \\
=&    \int  \Phi (G)  \left(  \cE_{\Om^c}^{ \ga,\mu}  \Phi(F)  \right)    \left[\frac{   d m_{  \ga, \la \mu} }{ d m_{  \ga,  \mu} }
\right]   d m_{  \ga, \mu}  \\
=&    \int  \Phi(G)  \left(  \cE_{\Om^c}^{ \ga, \mu} \Phi(F) \right)  d m_{  \ga, \la \mu}  \\
\end{split}
\end{equation} 
Here   in the third step  we have used that  $\la =1$  on   $\Om$  and  lemma  \ref{bam}  to conclude that    
$ [  d  m_{\ga, \la \mu}/d m_{\ga, \mu} ]$ 
 is  $\Sigma_{\Om^c}$-measurable.
Since  polynomials     $\Phi(G)$  are dense in    $L^2(Q , \Sigma_{\Om^c},    m_{ \ga, \la  \mu} )$
the second identity follows.

\subsection{amplitudes}

We  now return to the main problem and      consider a    sphere  $(\bbC_{\infty}, \ga)$   with standard    discs  $D_i$
 where  $\ga =  \ga_i$.  At    first  there is  no  parametrization  and   we just study the functions
  $< \prod_{i=1}^n   \Phi(F_i)     >_{\ga,  \mu  } $   with     $F_i  \in  \cS_{D_i}$.  
  
  First note that   $\Phi(F_1)$  is measurable with respect to  $\Sigma_{  D_1}$  and   $\Phi(F_2 ) \cdots  \Phi(F_n)$
is measurable with respect   to  $\Sigma_{   D'_1}$.  Thus by the Markov property  
\begin{equation}
\begin{split}
< \prod_{i=1}^n   \Phi(F_i))     >_{\ga,  \mu  } =  &
 <  \cE^{\ga,\mu}_{  D_1} \Phi(F_1)  \cE^{\ga,\mu}_{D'_1}\left( \Phi(F_2 ) \cdots  \Phi(F_n)\right)>_{\ga, \mu}\\
 =  &
 < ( \cE^{\ga,\mu}_{C_1} \Phi(F_1))\     \Phi(F_2 ) \cdots  \Phi(F_n)  >_{\ga, \mu}\\
\end{split}
\end{equation}
The same argument works with   $\cE^{\ga,\mu }_{C_{1++}}  \Phi(F_1)$ and also   by the same argument   we can successively replace  each $\Phi(F_i)$ by   
$\cE^{\ga,\mu }_{C_{i++}}  \Phi(F_i)$ 
Thus we have  
\begin{equation} 
  < \prod_{i=1}^n   \Phi(F_i)     >_{\ga,\mu}  
   =    < \prod_{i=1}^n  \cE^{\ga,\mu}_{C_{i++}}  \Phi(F_i)     >_{\ga,\mu}  
\end{equation}
By  Holder's inequality
\begin{equation}  \label{special}
|< \prod_{i=1}^n   \Phi(F_i)     >_{\ga,\mu}|   \leq  
    \prod_{i=1}^n   \| \cE^{\ga,\mu}_{C_{i++}} \Phi(F_i)\|_{n,\ga, \mu}
\end{equation}
where the norm is in   $ L^n(Q, \Sigma,   d m_{\ga,\mu})$.  
Thus we  study    the norms  $ \| \cE^{\ga,\mu}_{C_{i++}} \Phi(F_i)\|_{n,\ga, \mu}$.
We would like to   replace   the  metric  $\ga$ by  the standard   $\ga_i$ and  the $L^n$ norm 
by the $L^2$ norm.  \bigskip

Define   $\la_i$  by  $\ga =   \la_i  \ga_i$.
 Since  $\ga = \ga_i$  on   $D_{i++}$  
we have     $\la_i =1$  on   $D_{i++}$.

\begin{lem}   
Let   $F_i \in \cS_{D_i}$    and suppose  
\begin{equation}  \frac{1}{p}  <  \de  \equiv   \inf_{i,x}   \la_i(x)                                       
   \end{equation}
   Then   for each $n$ there is a constant  $C$ such that 
\begin{equation}  \label{robin}
|< \prod_{i=1}^n  \Phi( F_i)     >_{\ga,\mu}|   \leq   C   \prod_{i}    \|  \cE^{\ga_i,\mu}_{C_{i++}}\Phi(F_i) \|_{np,\ga_i,\mu} 
\end{equation}
\end{lem}
\bigskip

\pr    
By  (\ref{special})  this reduces to an estimate on  $  \| \cE^{\ga,\mu}_{C_{i++}}\Phi(F_i)\|_{n,\ga, \mu}  $. We compute  
with   
\begin{equation}
\De_i   =    \left[\frac{   dm_{  \ga_i, \la_i \mu} }{ dm_{  \ga_i,\mu } }
\right] 
\end{equation}
that  
\begin{equation}
\begin{split}
 \| \cE^{\ga,\mu}_{C_{i++}}\Phi(F_i)\|^n_{n,\ga, \mu} = &
    \int     |\cE^{\ga,\mu}_{C_{i++}}\Phi(F_i)|^n \  d  m_{\ga, \mu}   \\ 
     = &
    \int     |\cE^{\ga,\mu}_{C_{i++}}\Phi(F_i)|^n\   d  m_{\ga_i, \la_i \mu}  \hspace{1cm}    \textrm{        (by lemma  \ref{slam})     }   \\ 
      = &
    \int     |\cE^{\ga_i,\mu}_{C_{i++}}\Phi(F_i)|^n\   d  m_{\ga_i, \la_i \mu}  \hspace{1cm}    \textrm{        (by lemma  \ref{thanks})     }   \\ 
       = &
    \int     |\cE^{\ga_i,\mu}_{C_{i++}}\Phi(F_i)|^n  \  \De_i\  d  m_{\ga_i,  \mu}  \hspace{.9cm}    \textrm{        (by lemma  \ref{bam})     }   \\ 
  \leq    &
      \|  \left[\cE^{\ga_i,\mu}_{C_{i++}}\Phi(F_i)\right]^n\|_{p,\ga_i,\mu}  \|\De_i   \|_{q,\ga_i,\mu}    \\ 
  \end{split}
\end{equation}
where    $1/p + 1/q  =1$.
Equivalently 
\begin{equation}
 \| \cE^{\ga,\mu}_{C_{i++}}\Phi(F_i)\|_{n,\ga, \mu}   \leq   
   \|  \cE^{\ga_i,\mu}_{C_{i++}}\Phi(F_i)\|_{np,\ga_i,\mu}  \|\De_i   \|^{1/n}_{q,\ga_i,\mu}   
\end{equation}
Since   $1-1/q  <  \de   $    we have    $1-1/q <   \de_i  \equiv   \inf_x  \la_i(x)$, 
hence   $q <  (1-\de_i)^{-1}$,  and hence    the factor   $ \|\De_i   \|_{q,\ga_i,\mu}$  is finite by  lemma   \ref{bam}.
This completes the proof.
\bigskip

\res
\begin{enumerate}
\item   Hereafter we use the abreviated notation   $\cE^i_{A}  =  \cE^{\ga_i,\mu}_{A}$.

\item  For the next result we use a hypercontractivity estimate.
The general result is the following.    Let  $T$ be a bounded operator on a real Hilbert space  $H$  and suppose for   $s<t$    
\begin{equation}
   \|  T\|   \leq  \sqrt{ \frac{s-1}{t-1} }
\end{equation}
Let   $\phi(h)$ be associated Gaussian process on  $(Q, \Sigma,  m)$.
Then   $\Ga(T)$ defined as in   (\ref{second}) is a contraction from   $L^s(Q, \Sigma,  m) $
  to  $L^t(Q, \Sigma,  m)$, i.e.  
\begin{equation}
\|\Ga (T)  \psi  \| _t   \leq    \|    \psi   \|_s
\end{equation}
This result is due to Nelson   \cite{Nel73b},  \cite{Nel73c},  \cite{Si75}. 
\end{enumerate}
 \bigskip
  
\begin{lem}  \label{sugar}
Let   $F_i \in \cS_{ D_i+}$.  Then 
  for   $\mu$ sufficiently large
\begin{equation}
|< \prod_{i=1}^n   \Phi(F_i)     >_{\ga,\mu}|   \leq   C    \prod_{i}    \|  \cE^{i}_{C_{i+}}\Phi(F_i) \|_{2,\ga_i,\mu} 
\end{equation}
\end{lem}
\bigskip

\pr 
Applying the Markov property twice
we   have 
\begin{equation}  \label{styx}
\cE^i_{C_{i++}}\Phi(F_i) =    \cE^i_{D'_{i++}}\Phi(F_i)
 =      \cE^i_{D'_{i++}}\cE^i_{C_{i+}}\Phi(F_i) =  
   \cE^i_{D'_{i++}} \cE^i_{D_{i+}} \cE^i_{C_{i+}}\Phi(F_i)
\end{equation}
 Thus     from  (\ref{robin})
\begin{equation}
\begin{split}
|< \prod_{i=1}^n   \Phi(F_i)     >_{\ga,\mu}|   \leq   & C  \prod_{i}  
 \|  \cE^i_{D'_{i++}} \cE^i_{D_{i+}} \cE^i_{C_{i+}}\Phi(F_i)
\|_{np,\ga_i, \mu}
 \\
  =   & C   \prod_{i}  
   \|\Ga(  e^i_{D'_{i++}} e^i_{D_{i+}} )  \cE^i_{C_{i+}}\Phi(F_i)\|_{np,\ga_i, \mu}
 \\
\end{split}
\end{equation}
In a following lemma we show that as  $\mu \to \infty$
\begin{equation}  \label{dinky}
 \| e^i_{D'_{i++}} e^i_{D_{i+}} \|  \leq   \cO(\mu^{-1/2 + \ep})
\end{equation}
For      $\mu $ sufficiently large  (depending on  $\ga, n$) this implies that  
\begin{equation}
 \|e^i_{D'_{i++}} e^i_{D_{i+}} \|   \leq      \sqrt{  \frac{1}{np-1} }
\end{equation}
Then by  the    hypercontractive bound with  $s=2$  and  $t = np$  we conclude that   
\begin{equation}
  \|\Ga(  e^i_{D'_{i++}} e^i_{D_{i+}} )  \cE^i_{C_{i+}}\Phi(F_i)\|_{np,\ga_i, \mu}
\leq      \|  \cE^i_{C_{i+}}\Phi(F_i)\|_{2,\ga_i, \mu}
\end{equation}
whence the result.
\bigskip

\begin{lem}   \label{soup}
Let   $(\bbC_{\infty}, \ga)$ be the sphere with conformal metric and let 
$\La_1,  \La_2 $  be disjoint closed subsets.  Then in the Sobolev 
space   $H^{-1}_{\ga, \mu}$
\begin{enumerate}
\item  For any  $\ep >0$  we have as    $\mu \to \infty$
\begin{equation}
   \|e_{\La_1} e_{\La_2}\|_{-1}  =   \cO(\mu^{-1/2+ \ep})
\end{equation}
\item 
$e_{\La_1} e_{\La_2}$ is Hilbert Schmidt
\end{enumerate}
\end{lem}
\bigskip

\re   Similar results  are known on $\bbR^2$,   see  Simon  \cite{Si75} who attributes
the    idea to E. Stein.  Our proof is a straightforward    adaptation to the sphere.   \bigskip

\pr The norm in   $H^{+1}_{\ga, \mu}$
can be written   in terms of the exterior derivative   and the  $L^2(\bbC^{\infty}, \mu_{\ga})$  norm as
   \begin{equation}
\|f \|^2_{+1}  =  \|df\|^2   +  \mu \|f\|^2
\end{equation}
Choose a  smooth function $g$ so that   $g=1$ on  $\La_1$
and  $g=  -1$ on  $\La_2$  and  $\|g\|_{\infty}   =1$.  Then   for  
$\al>0$ 
\begin{equation}
\begin{split}
\| g  f \|^2_{+1}   =&  \|  d (gf) \|^2   +  \mu   \|gf \|^2  \\
\leq   & \left(  \|df\|  +  \|dg\|_{\infty} \|f\| \right)^2      +  \mu   \|f\|^2 \\
 \leq &  (1 + \al^{-1})   \|df\|^2    +     (1 + \al)    \|dg\|^2_{\infty} \|f\| ^2  +  \mu \|f \|^2    \\
 \leq &  (1 + \al^{-1})   \|f\|^2_{+1}   \\
\end{split}
\end{equation}
where the last step holds provided     $   (1 + \al)    \|dg\|^2_{\infty}  \leq   \mu  \al^{-1}$ .
Now choose   $\al  =  \mu^{1/2-\ep}$   and   $\mu$  sufficiently 
large so the inequality holds.
\footnote{ For  $\mu \geq 1$ it suffices that   $\|dg\|_{\infty} \leq  \mu^{2\ep}$.   Note that the larger the distance between the sets,  the smaller one can take
$\|dg\|_{\infty} $ and hence the weaker the restriction  on  $\mu$.}
  Then   with  $ \beta^2  =  1+\al^{-1}  =  1 + \mu^{-1/2+ \ep}$
have  
\begin{equation}\| g  f \|_{+1}     \leq    \beta   \|f\|_{+1}
\end{equation}

Referring to the  $H^{+1},  H^{-1}$ pairing the dual operator 
  to multiplication by $g$  on   $H^{+1}$ is multiplication by  $g$  in 
  $H^{-1}$.   It has the same norm and so 
\begin{equation}
\| g  f \|_{-1} \leq  
  \beta    \|f \|_{-1}    
  \end{equation}

Now suppose  $\supp\  f  \subset \La_1$ and  $\supp\    h  \subset  \La_2$ so
that   $f+h =  g(f-h)$.  Then we have
\begin{equation}
4(f,h)_{-1}  =  \|f+h\|^2_{-1}-   \|f-h\|^2_{-1}   
\leq    (\beta -1)    \|f-h\|^2_{-1} 
\end{equation}
Expanding    $\|f-h\|^2_{-1}  =  \|f\|^2_{-1} -   2 (f,h)_{-1}   +\|h\|^2_{-1}$
we can rewrite this as 
\begin{equation}  
(f,h)_{-1}  \leq  \left( \frac{\beta-1}{\beta+1}   \right)   \left( \frac{ \|f\|^2_{-1} + \|h\|^2_{-1}}{2} \right)
\end{equation}
Replacing  $f$  by  $f/ \|f\|_{-1}$, etc.  we obtain  the same bound but with    $  \|f\|_{-1}  \|h\|_{-1}$
on the right.   Replacing  $f$  by  $-f$  we get the same bound with a minus sign on the left. 
Hence
still with   $\supp\  f  \subset \La_1$ and  $\supp\    h  \subset  \La_2$  we have  
\begin{equation}  
|(f,h)_{-1}|  \leq   \left( \frac{\beta-1}{\beta+1}   \right)     \|f\|_{-1}  \|h\|_{-1}
\end{equation}
Since   $\|    e_{\La}   \|  \leq  1$ we have   for any   $f,h$  
\begin{equation}  
|(e_{\La_1}f,e_{\La_2}h)_{-1}|  \leq   \left( \frac{\beta-1}{\beta+1}   \right)     \|f\|_{-1}  \|h\|_{-1}
\end{equation}
and so   
\begin{equation}
\|e_{\La_1}e_{\La_2} \|_{-1} \leq     \frac{\beta-1}{\beta+1}  \leq      \frac{\beta^2-1}{\beta^2+1}   
 =  \frac{   \mu^{-1/2+ \ep}  }{  2+  \mu^{-1/2+ \ep} }  
\end{equation}
This proves the first result.

For the second result    let  $\om  =   \sqrt  {  - \De_{\ga}   +   \mu  }$ regarded as  a unitary 
from  from $H^{+1}$ to $L^2$ or from   $L^2$ to  $H^{-1}$.    Let  
 $\zeta_1$   be  smooth  and equal to    $1$  on  $\La_1$,   let   $\zeta_2$
  be   smooth  and  equal to    $1$  on  $\La_2$, and let  $\zeta_1 \zeta_2 =0$.  Then we have
\begin{equation}
\begin{split}
&(e_{\La_1}f,  e_{\La_2}h)_{-1} =     (\zeta_1e_{\La_1}f, \zeta_2 e_{\La_2}h)_{-1} 
= (\om^{-1}\zeta_1e_{\La_1}f, \om^{-1}  \zeta_2 e_{\La_2}h)_{\ga}
= ( \om^{-1}e_{\La_1}f,   A  \    \om^{-1}   e_{\La_2}h)_{\ga}
\end{split}
\end{equation}
where   $A$ is the operator on  $L^2(\bbC^{\infty}, \mu_{\ga})$
\begin{equation}
A  = \om  \zeta_1  \om^{-2}  \zeta_2  \om
\end{equation}
Therefore
\begin{equation}
e_{\La_1, \La_2}    = [ \om^{-1}   e_{\La_1}] ^*A  [\om^{-1}   e_{\La_2}]
\end{equation}
But   $ [\om^{-1}   e_{\La_2}] $  is bounded and so  is  
$ [ \om^{-1}   e_{\La_1}] ^*= [   e_{\La_1}  \om] $.
Thus   it suffices to show  that $A$ is Hilbert-Schmidt.

Next   note that $ [- \De_{\ga} , \zeta]  =B$
 where 
  \begin{equation}
 (Bf)(x) =   (- \De_{\ga} \zeta)(x)f(x)  -  2(d \zeta(x), d f(x))_{\ga}
 \end{equation}
Using this and   $\zeta_1 \zeta_2 =0$  we can rewrite  
$A$ as 
\begin{equation}
A  = - [ \om^{-1}B ]\om^{-2} [B\om^{-1}]
\end{equation}
But   $\|B f  \|  \leq  \const  \|f\|_{+1}$.  Hence   $B\om^{-1}$  is bounded on $L^2$
and  so is the adjoint  $ \om^{-1} B$.  Finally
since  $\om^{-2} = ( -\De_{\ga} + \mu)^{-1}$ is Hilbert-Schmidt as noted earlier, 
we conclude that  $A$ is Hilbert-Schmidt to complete the proof.
\bigskip

To state our first main result we  create  Hilbert spaces    based on  the 
discs     $D_i$  analagous to the construction   of   section  \ref{nuts}  on   $D_0$.
  Consider   $\cS_{D_i}$  with the   norm 
$\|F\|^2  =  <  (\Phi( \Theta_i F) \Phi(F)>_{\ga_i, \mu}$
where   $\Theta_i$ is the map induced by   radial  reflection through  $C_i$.      Let  $\cN_i$ be the null space,   form 
the quotient space      $\cH_{i,0}  =   \cS_{D_i}/ \cN_i$
and   then take the completion
\begin{equation}  
\cH_i  =  \overline{ \cH_{i,0}}  =     \overline{  \cS_{D_i}/ \cN_i}
\end{equation}
If    $(\cdot,  \cdot)_i$ is the inner product in  $\cH_i$  and       $\nu_i$   maps elements
of    $\cS_{D_i}$  to   equivalence classes in   $ \cH_i$  then 
\begin{equation}
(\nu_i ( F ),    \nu_i (F'))_i  =  < \Phi( \Theta_i F) \Phi(F')>_{\ga_i, \mu}
\end{equation}
There  are also alternate constructions of $\cH_i$ analagous to the second and
third constructions in section \ref{nuts}.  In the third construction $\cH_i = 
 L^2( Q, \Sigma_{C_i},  m_{\ga_i,\mu}) $.
\bigskip

\begin{thm}  Let   $(\bbC_{\infty}, \ga)$  be a sphere  with standard discs   $\{D_i \}$. 
For   $F_i  \in \cS_{D_i}$  and $\mu$ sufficiently large:
\begin{enumerate}
\item    The expectation    $< \prod_{i=1}^n   \Phi(F_i)     >_{\ga,\mu }  $
depends  on $F_i$  only through  the equivalence class in  $\cH_{i,0}$   and
satisfies
\begin{equation}  \label{pie}
|< \prod_{i=1}^n   \Phi(F_i)     >_{\ga,\mu}|    \leq   C   \prod_{i=1}^n
\| \nu_i (F_i)  \|_i
\end{equation}
Hence it extends to  a bounded multilinear functional  on  $\cH_1  \times  \cdots  \times  \cH_n$.
\item   Given a constant  $Z_{\ga, \mu}$
there is a unique linear functional   
$A_{\ga, \mu} :   \cH_1  \otimes  \cdots  \otimes  \cH_n
 \to \bbC$
 such that  if  $\cF_i  =  \nu_i (F_i)  \in  \cH_i$ then  
 \begin{equation}  \label{sunday}
A_{\ga, \mu} (  \cF_1  \otimes  \cdots  \otimes  \cF_n  )  
=  Z_{\ga, \mu}< \prod_{i=1}^n   \Phi(F_i)     >_{\ga,\mu} 
\end{equation}
\end{enumerate}
\end{thm}
\bigskip

\pr   
Again by the Markov property twice
we   have 
\begin{equation}
\cE^i_{C_{i+}}\Phi(F_i) =    \cE^i_{D'_{i+}}\Phi(F_i)
 =      \cE^i_{D'_{i+}}\cE^i_{C_{i}}\Phi(F_i) =  
   \cE^i_{D'_{i+}} \cE^i_{D_i} \cE^i_{C_{i}}\Phi(F_i)=\cT_i\cE^i_{C_{i}}\Phi(F_i)
\end{equation}
where     $\cT_i =  \cE^i_{D'_{i+}} \cE^i_{D_i} $. 
Thus  lemma   \ref{sugar}  can be rewritten as
\begin{equation}  \label{chicken}
|< \prod_{i=1}^n   \Phi(F_i)     >_{\ga,\mu}|  
 \leq    C   \prod^n_{i=1}   \|\cT_i  \cE^{i}_{C_i} \Phi(F_i)\|_{2,\ga_i, \mu}
\end{equation}
At  first   we ignore the $\cT_i$  using   $\|\cT_i \|  \leq  1$  to write  
\begin{equation}  \label{goose}
|< \prod_{i=1}^n   \Phi(F_i)     >_{\ga, \mu}|  
 \leq    C   \prod^n_{i=1}   \| \cE^{i}_{C_i} \Phi(F_i)\|_{2,\ga_i, \mu}
\end{equation}
Just   as in   (\ref{sing})  we have   
\begin{equation}
 \| \cE^{i}_{C_i} \Phi(F_i)\|_{2,\ga_i, \mu}  =  \| \nu_i(F_i)  \|_i
\end{equation}
which establishes   (\ref{pie})  and gives the first result.

 For the second point we must show that  the multilinear  functional is a 
Hilbert-Schmidt functional, then the existence of the map is the universal property of 
the tensor product  (See \cite{KaRi83},  Theorem 2.6.4). 
Thus   if   $\ell( \cF_1, \dots,  \cF_n)   = Z_{\ga, \mu}   < \prod_{i=1}^n   \Phi(F_i)     >_{\ga,\mu } $   and
if   $\{  \Phi^i_{\al_i}  \}$ is an orthonormal 
basis in  $\cH_i$   we need
\begin{equation}
\sum_{\al_1, \dots,  \al_n}  | \ell(\Phi^1_{\al_1} ,   \dots,   \Phi^n_{\al_n})|^2
< \infty
\end{equation}
 Now   the unitary   identification  
$ \cE^i_{C_i} \Phi(F_i)  \leftrightarrow   \nu_i(F_i)  = \cF_i$  between the third and first constructions 
of  $\cH_i$    takes  $\cT_i:   L^2( Q, \Sigma_{C_i},  m_{\ga_i,\mu}) \to   L^2( Q, \Sigma,  m_{\ga_i,\mu})    $ 
to some  $\hat  \cT_i: \cH_i  \to   L^2( Q, \Sigma,  m_{\ga_i,\mu})    $.
By  (\ref{chicken})     our linear functional   satisfies\begin{equation}
 | \ell(\cF_1 ,   \dots,  \cF_n) |   \leq     C'   \prod^n_{i=1}   \|\hat  \cT_i  \cF_i\|
 \end{equation}
Thus we must show that  $  \hat  \cT_i$   or  $\cT_i$  is Hilbert-Schmidt.  

We   have
\begin{equation}
   \cT_i =  \cE^i_{D'_{i+}} \cE^i_{D_i} = \Ga( e^i_{D'_{i+}} e^i_{D_i} )
\equiv    \Ga  (T_i)
\end{equation}
Since $D'_{i+}$   and   $D_i$ are disjoint 
we    know that  $T_i$ is Hilbert-Schmidt   by   lemma  \ref{soup}.     Hence     $T^*_iT_i$ is 
trace class.   By the next lemma      $\Ga (T^*_iT_i)=  \Ga(T_i)^*   \Ga  (T_i)$ is 
trace class.   Then    $\cT_i =  \Ga(T_i)$   is   Hilbert-Schmidt on    $L^2( Q, \Sigma,  m_{\ga_i,\mu}) $
and hence so is the restriction to  the subspace   $L^2( Q, \Sigma_{C_i},  m_{\ga_i,\mu}) $.
\bigskip

\begin{lem} Let  $H$ be a real Hilbert space and let   $\phi(f)$ be the Gaussian process indexed
by   $H$ on  $(Q, \Sigma, m)$.   
If    $T \geq 0$   is a   trace class  contraction    on    $\cH$,  then   $\Ga(T)\geq 0 $ is a   trace class contraction  on  $L^2(Q, \Sigma, m)$.
\end{lem}
\bigskip
                   
\re  This lemma is familiar from  the proof   that  the partition function for the free boson gas
is finite.
\bigskip
  
\pr   Let  $e_k$ be a basis of eigenfunctions for  $T$    with  $T  e_k  =  \la_k e_k$ and   $0 \leq \la_k <1$ and 
$\sum_k  \la_k  < \infty$.
Then there is an associated basis for the $L^2$ space  indexed by finite   sequences 
of non-negative integers $\{ n_k\}_{k=1}^N$
and given by 
\begin{equation}
\Phi_{ \{n_k\} }  = \prod_k     \frac{1}{\sqrt{n_k!}} :   \phi(e_1^{n_1})  \cdots   \phi(e_N^{n_N}):
\end{equation}
We have  
\begin{equation}
\Ga (T)  \Phi_{ \{n_k\} }   = \left( \prod_{k} \la_k^{n_k} \right) \Phi_{ \{n_k\} }  
\end{equation}
Now we can compute
\begin{equation}
 \textrm{Tr } (\Ga(T))  = \sum_{ \{n_k\} }  \prod_{k}   \la_k^{n_k}
=  \prod_k  \sum_n  \la_k^{n}  =  \prod_k  \frac{1}{1-\la_k}
\end{equation}
The product converges since  $\log(1-\la_k)  =  \cO(\la_k)$  and  $\sum_k  \la_k  < \infty$.
This completes the proof of the lemma and the theorem.
\bigskip

Now suppose we have  a   parametrized    sphere   as  explained in section  \ref{bunko}.
Then we can refer everything to our standard Hilbert space $\cH$  based on  $D_0$.

\begin{thm}   \label{elmer}
 Let   $(\bbC_{\infty}, \ga)$  be  
 a  parametrized sphere with in-discs
  $\{ D_i \}_{i \in I}$   and out-discs   $\{  D_i \}_{i \in I'}$.   Let   $\mu$ be sufficiently large.
 Then    there is a unique operator   
\begin{equation}  
A^{I'I}_{\ga, \mu }:   \otimes_{i \in I}  \cH \  \to\  \otimes_{i \in I'} \cH  
\end{equation}
such that if   $F_i  \in   \cS_{D_0}$ and 
$\cF_i  =  \nu(F_i)  \in \cH  $  then
\begin{equation}   \label{bucket}
\left( [\otimes_{i \in I'}  \cF_i],   A^{I'I}_{\ga,\mu}  [\otimes_{i \in I} \cF_i ]\right) = Z_{\ga, \mu}
< \prod_{i\in I'}  \Phi(\cJ'_i\Theta F_i) \prod_{i \in I}   \Phi(\cJ_i F_i)   >_{\ga,\mu}
\end{equation}
Furthermore  $A^{I'I}_{\ga, \mu}$ is Hilbert-Schmidt.
\end{thm}
\bigskip

\pr 
The map  $j_i$ defined in (\ref{soon})  satsifies  $j_i^* \ga_0 = \ga_i$  and so the pullback
$j_i^*$ on  smooth    functions    extends to a unitary from   $L^2(\bbC_{\infty}, \mu_{\ga_0}) $ to   $L^2(\bbC_{\infty}, \mu_{\ga_i}) $.  We also have   $j_i^* \De_{\ga_0}  =  \De_{\ga_i} j_i^*$  and so  
$j_i^*$ also determines a unitary map from  $H^{-1}_{\ga_0, \mu}$  to   $H^{-1}_{\ga_i, \mu}$
which takes elements with support in  $D_0$  to elements with support in  $D_i$.  
Furthermore  $\theta j_i  =  j_i \theta_i$  hence    $j_i^*\theta^* =  \theta_i^* j^*_i$   and combining 
these facts   
\begin{equation}
<   \Phi(\Theta F) \Phi(F)  >_{\ga_0, \mu}   =
<   \Phi( \cJ_i \Theta  F) \Phi(\cJ_i F)  >_{\ga_i, \mu}  =<   \Phi(\Theta_i \cJ_iF) \Phi(\cJ_i F)  >_{\ga_i, \mu}  
 \end{equation}
Thus the   map   $\cJ_i: \cS_{D_0} \to   \cS_{D_i}$  is norm preserving and so   determines a unitary
$U_i:  \cH \to  \cH_i$ such that  
\begin{equation}
U_i\  \nu(  F)   =  \nu_i ( \cJ_i F)
\end{equation}
The operator   $(j'_i)^*$  is also unitary   from  $H^{-1}_{\ga_0, \mu}$  to   $H^{-1}_{\ga_i, \mu}$
and  $(j'_i)^*\theta^* =  \theta_i^*( j'_i)^*$   and so  
\begin{equation}
<   \Phi(\Theta F) \Phi(F)  >_{\ga_0, \mu}   =
<   \Phi( \cJ'_i \Theta  F) \Phi(\cJ'_i F)  >_{\ga_i, \mu}  =<   \Phi( \cJ'_i \Theta F) \Phi(\Theta_i \cJ'_i\Theta F)  >_{\ga_i, \mu}  
 \end{equation}
Thus the    map   $  \cJ'_i \Theta: \cS_{D_0} \to   \cS_{D_i}$  is norm preserving and determines an   anti-unitary
$V_i:  \cH \to  \cH_i$ such that  
\begin{equation}
V_i\  \nu(  F)   =  \nu_i ( \cJ'_i  \Theta     F)
\end{equation}

Let $\cH_I =   \otimes_{i\in I}  \cH$
and    $\cH_{I'}  =  \otimes_{i \in I'} \cH $.
We define  first a bounded     linear functional  
$A^{I'I}_{\ga, \mu}:  \cH_{I'}    \otimes  \cH_I
\to  \bbC$
(anti-linear in $\cH_{I'}$)   by 
\begin{equation}
A^{I'I}_{\ga, \mu}  =  A^{I'I}_{\ga, \mu} \circ  
 \left(   \left( \otimes_{i\in I'} V_i \right)  \otimes
  \left(  \otimes_{i \in I} U_i   \right)  \right)
\end{equation}
Then  by  (\ref{sunday})
\begin{equation}  \label{run}
\begin{split}
 A^{I'I}_{\ga,\mu}\left( [\otimes_{i \in I'}  \cF_i]  
 \otimes    [\otimes_{i \in I} \cF_i ]\right) \
=&  A_{\ga, \mu} \left( [\otimes_{i \in I'} V_i \cF_i]  
\otimes    [\otimes_{i \in I}U_i \cF_i ]\right) \\
=&  A_{\ga, \mu} \left( [\otimes_{i \in I'} \nu_i ( \cJ'_i \Theta F)] 
 \otimes    [\otimes_{i \in I} \nu_i( \cJ_i F) ]\right) \\
=&Z_{\ga, \mu} < \prod_{i\in I'}   \Phi(\cJ'_i\Theta F_i) \prod_{i\in I}   \Phi(\cJ_i F_i)   >_{\ga,\mu}\\
\end{split}
\end{equation}
If   $\Phi_{\al}$ is an orthonormal basis for  $\cH_{I}$ and   
 $\Phi'_{\beta}$ is an orthonormal basis for  $\cH_{I'}$  then  $\Phi'_{\beta}  \otimes \Phi_{\al}$
is an orthonormal basis for   $\cH_{I'} \otimes  \cH_{I}$.  Since   $A^{I'I}_{\ga, \mu}$ is 
a bounded linear functional on this space
  \begin{equation}   \label{om}
\sum_{\al,\beta} | A^{I'I}_{\ga, \mu} (  \Phi'_{\beta}  \otimes   \Phi_{\al}  )|^2     < \infty 
\end{equation}
The  bounded   linear functional  $ A^{I'I}_{\ga, \mu} $   determines   a    bounded    bilinear form
 $ A^{I'I}_{\ga, \mu} $  on  $\cH_{I'} \times  \cH_I$ (anti-linear in $\cH_{I'}$)   such that   $ A^{I'I}_{\ga, \mu} (\Phi'  \otimes  \Phi)=A^{I'I}_{\ga, \mu} (\Phi', \Phi) $.  The bilinear form   determines a bounded operator  $A^{I'I}_{\ga, \mu} $  from  $\cH_I$ to $\cH_{I'}$
such that   $A^{I'I}_{\ga, \mu} (\Phi', \Phi)=(\Phi',A^{I'I}_{\ga, \mu}  \Phi)$.    Then   (\ref{run}) says that the operator
satisfies   (\ref{bucket})   and   
(\ref{om})   says that the operator is Hilbert-Schmidt.

\subsection{sewing}

We now establish a sewing property in a simple configuration.  
This is facilitated by a special choice of the constant  $Z_{\ga, \mu}$.
  If   $\ga = \la \ga_0$ we take   
\begin{equation}
Z_{\ga, \mu}  = \int   \exp\left( - \frac12  : \phi^2: (\la \mu - \mu) \right) \  d m_{\ga_0,\mu}
\end{equation}

  We start 
by finding   a more explicit representation of the operators $ A^{I'I}_{\ga,\mu}$  
in the case where  there is one out-disc.
Consider a parametrized sphere   $(\bbC_{\infty}, \ga)$   with    out-disc   $D'_0$   (with the identity 
parametrization)       and   
in-discs   $\{D_i\}_{i \in I}$         in    $D_0$.    Then        
   $\ga  = \ga_0$  on a neighborhood of    $D'_0$ and 
  so if      $\ga =  \la  \ga_0$  then  $\la=1$  on a neighborhood of    $D'_0$.   We consider the
  corresponding   amplitude  denoted  $A^{1I}_{\ga,\mu} $.
With     $\cF  =  \nu(F),   \cF_j  = \nu(F_j)$
we compute
  \begin{equation} 
\begin{split}
\left(  \cF,   A^{1I}_{\ga_,\mu}  [\otimes_{i \in I} \cF_I ]\right)
 =& Z_{\ga, \mu}   \int  \Phi(\Theta  F)   \prod_{i \in I}\Phi(\cJ_iF_i)   d m_{\ga,\mu}\\
 =& Z_{\ga, \mu}    \int  \Phi(\Theta F)   \prod_{i \in I} \Phi((\cJ_iF_i)_{\la})   d m_{\ga_0,\la\mu}\\
 =& Z_{\ga, \mu}   \int  \Phi(  \Theta F)   \prod_{i \in I}\Phi((\cJ_iF_i)_{\la})   \left[\frac{   d m_{  \ga_{0}, \la \mu} }{ d m_{  \ga_{0},\mu} }     \right]  d m_{\ga_0,\mu}\\
  =&   \int  \Phi(  \Theta F)   \prod_{i \in I}\Phi((\cJ_iF_i)_{\la})  e^{ -  : \phi^2: (\la \mu - \mu)/2} \ 
 d m_{\ga_0,\mu}\\
  =&   \left(\nu ( \Phi(  F)) ,
  \nu
  \left (  \prod_{i \in I}\Phi((\cJ_i   F_i)_{\la})  e^{ -  : \phi^2: (\la \mu - \mu)/2}    \right)  \    \right)   \\
 \end{split}
\end{equation}
Here we have used  lemma \ref{slam}  and  lemma \ref{bam}.   In the fourth step we  
use the constant      $Z_{\ga,\mu}$ to cancel the denominator  in   the expression (\ref{RN}) for 
  $  [ d m_{  \ga_{0}, \la \mu} /d m_{  \ga_{0},\mu}  ] $.   In the last step we  take 
$ \Phi(  \Theta F)  =  \Theta (\Phi(F))$  and    use the 
second construction of  $\cH$,  taking into account that  
   $\la =1$  on  $D'_0$
so $ \exp( -  : \phi^2: (\la \mu - \mu)/2)$  is measurable with respect to  $\Sigma_{D_0}$ by lemma \ref{bam}.    Since   $\nu( \Phi(F))   \leftrightarrow  \nu(F)= \cF$ under the equivalence of constructions 
 we conclude that 
\begin{equation}   \label{ice}
 A^{1I}_{\ga_,\mu}  [\otimes_{i \in I} \cF_j ]
=    \nu
  \left(  \prod_{i \in I}\Phi((\cJ_i   F_i)_{\la}) e^{ -  : \phi^2: (\la \mu - \mu)/2}  
      \right) 
\end{equation}

Also consider a parametrized sphere   $(\bbC_{\infty}, \ga)$  with in-disc   $D_0$ (with the identity 
parametrization)  and 
out-discs   $\{ D_i \} $   in $ D'_0$.  Then   $\ga  = \ga_0$  on a neighborhood of     $D_0$ and   so if     $\ga =  \la  \ga_0$ then  $\la=1$ 
on a neighborhood of  $D_0$.   We consider the corresponding   amplitude  denoted
 $A^{I1}_{\ga,\mu} $.  Then 
   \begin{equation}  
\begin{split}
\left(   [\otimes_{i \in I'} \cF_i ],   A^{I'1}_{\ga,\mu}  \cF  \right) 
=&  Z_{\ga, \mu}   \int      \left (   \prod_{i \in I'} \Phi(\cJ'_i \Theta F_i)\right)   \Phi(F)  \  d m_{\ga,\mu}\\
=& Z_{\ga, \mu}    \int      \left (   \prod_{i \in I'} \Phi((\cJ'_i \Theta   F_i)_{\la})\right)   \Phi(F) \   d m_{\ga_0,  \la\mu}\\
 =&   Z_{\ga, \mu}  \int      \left (   \prod_{i \in I'}\Phi((\cJ'_i \Theta F_i)_{\la})\right)   \Phi(F) \ 
   \left[\frac{   d m_{  \ga_{0}, \la \mu} }{ d m_{  \ga_{0},\mu} }\right]   \     d m_{\ga_0,\mu}\\
 =&   \int      \left (   \prod_{i \in I'}  \Phi((\cJ'_i \Theta F_i)_{\la})\right)   \Phi(F) \ 
 e^{ -  : \phi^2: (\la \mu - \mu)/2}    \     d m_{\ga_0,\mu}\\
  =&   \left( \nu  \left ( \Theta     e^{ -  : \phi^2: (\la \mu - \mu)/2}   \prod_{i \in I'} \Phi((\cJ'_i \Theta F_i)_{\la})\right) ,  \nu( \Phi(F) ) \  \right) \\
\end{split}
\end{equation}
Then we have  
       \begin{equation}   \label{cream}
(A^{I,1}_{\ga,\mu}  )^*   
 [\otimes_{i \in I'} \cF_i ]   =  \nu  \left ( \Theta     e^{ -  : \phi^2: (\la \mu - \mu)/2}   \prod_{i \in I'} \Phi((\cJ'_i \Theta F_i)_{\la})\right)\end{equation}
\bigskip

We sew together  two   such   amplitudes  by  composition.
Here is  the result for   a    simple configuration:

\begin{thm}  \label{over}
Let   $(\bbC_{\infty},  \ga )$ be a parametrized sphere with in-discs   $\{ D_i \}_{i \in I}$
contained  in  $D_0$ 
and  out-discs  $\{ D_i \}_{i \in I'}$    contained  in   $D'_0$.
Let  $\ga_1$ be a metric with  $\ga_1 = \ga$ on $D_0$  and  $\ga_1 =  \ga_0$ on  $D'_0$.
Let  $\ga_2$ be a metric with  $\ga_2 = \ga_0$ on $D_0$  and  $\ga_2 =  \ga$ on  $D'_0$.
Then
\begin{equation}    \label{snow}
 A^{I'1}_{\ga_2,\mu}  A^{1I}_{\ga_1,\mu}   =   A^{I'I}_{\ga,\mu}
 \end{equation}
 \end{thm}
 
 \pr  We   have   $\ga_1 = \la_1 \ga_0,\  \ga_2 = \la_2 \ga_0,\  \ga =  \la \ga_0$.
Then     $\la_1=  \la$  on  $D_0$  and      $\la_1 =  1$  on  $D'_0$.  Also    
  $\la_2=  1$   on $D_0$  and  $\la_2 =  \la$  on  $D'_0$.
Then we compute using  (\ref{ice}) and (\ref{cream})
\begin{equation}
\begin{split}
& \left(  [\otimes_{i \in I'} \cF_i ],   A^{I'1}_{\ga_2,\mu}  A^{1I}_{\ga_1,\mu}  [\otimes_{i \in I} \cF_j ]\right)\\
=&  \left( \nu  \left ( \Theta     e^{ -  : \phi^2: (\la_2 \mu - \mu)/2}   \prod_{i \in I'} \Phi((\cJ'_i \Theta F_i)_{\la_2})\right),  
      \nu  \left(  \prod_{i \in I}\Phi((\cJ_i   F_i)_{\la_1}) e^{ -  : \phi^2: (\la_1 \mu - \mu)/2}   \right) 
   \right) \\
   =&  \int       \prod_{i\in I'}\Phi((\cJ'_i\Theta  F_i)_{\la})    
          \prod_{i \in I}\Phi((\cJ_iF_i)_{\la})  e^{ -  : \phi^2: (\la \mu - \mu)/2}    \ \   d m_{\ga_0,\mu}  \\
  =& Z_{\ga,\mu}   \int       \prod_{i \in I'}\Phi((\cJ'_i\Theta  F_i)_{\la})    
          \prod_{i \in I}\Phi((\cJ_iF_i)_{\la})    \left[\frac{   d m_{  \ga_{0}, \la \mu} }{ d m_{  \ga_{0},\mu} }\right]    \ \   d m_{\ga_0,\mu}  \\
 =&  Z_{\ga,\mu}  \int       \prod_{i \in I'}\Phi((\cJ'_i\Theta     F_i)_{\la})      
  \prod_{i \in I}\Phi((\cJ_iF_i)_{\la})     \ \      d m_{\ga_0,\la  \mu}  \\
   =&  Z_{\ga,\mu}  \int       \prod_{i \in I'}\Phi(\cJ'_i\Theta  F_i)   
          \prod_{i \in I}\Phi(\cJ_iF_i )   \ \   dm_{\ga,  \mu}  \\
=& \left(  [\otimes_{i \in I'} \cF_i ],   A^{I'I}_{\ga,\mu}  [\otimes_{i \in I} \cF_i ]\right)\\
\end{split}
\end{equation}
Here we have used   $
(\la_1 -1 )  +   (\la_2 -1)  =  \la -1$  in the second step.  This completes the proof.
\bigskip

Since the product of two Hilbert-Schmidt operators is trace class we have 

\begin{cor}
$ A^{I'I}_{\ga,\mu}$  is trace class
\end{cor}

\res
\begin{enumerate}
\item   The fact that we are sewing together   the   out-disc  $D'_0$  with
a   the      in-disc  $D_0$  was just a convienience.    More general configurations
can be treated by the same methods.

\item Because all our amplitudes refer to spheres    we have
managed to avoid any  actual  sewing of manifolds
   A  somewhat   different approach to sewing  was developed in  \cite{Dim04}.   It  had the
advantage of working for any  compact   Riemann  surface,   but  the disadvantage that
the identities like  (\ref{snow}) did not hold.

\item   Another way to characterize our sewing theorem is to take  an orthonormal
basis  $\Phi_{\al}$   for  $\cH$ 
and   then  with   $\Psi' =   \otimes_{i \in I'} \cF_i $   and  $\Psi =   \otimes_{i \in I} \cF_i$
we have  
\begin{equation}
\sum_{\al}( \Psi,  \ A^{I'1}_{\ga_2,\mu} \Phi_{\al})(  \Phi_{\al}, A^{1I}_{\ga_1,\mu} \Psi) 
 =  ( \Psi',  A^{I'I}_{\ga,\mu}   \Psi)
 \end{equation}
 In the same way we     could sew together two legs on the same sphere    
forming    for example
\begin{equation}
\sum_{\al}   ( (\Psi  \otimes  \Phi_{\al})  ,   A^{I'  \cup1,I  \cup 1}_{\ga,\mu}  
(\Phi_{\al}    \otimes  \Psi' ) )
\end{equation}  
Our estimates are good enough to show that the sum converges.   However
this should be identified   with   an  amplitude  on  a  torus    which 
is outside the scope of this paper.
\end{enumerate}

\section{Massless fields}
      
 \subsection{fields}

Now consider  massless fields.   In this case    we have a conformal field theory which we develop following
Gawedski    \cite{Gaw99}. The interesting fields are now the exponential fields 
$ e^{ik\phi(x)}$.  If  we restrict  to integer     $k$  the these are the fields  that  that  
 occur in circle valued    compactification.

The fields    $ e^{ik\phi(x)}$ are   singular objects  and we will 
need a regularized version  denoted  $ [ e^{ik\phi(x)}]_r$.    Our first task  is to give a meaning to expressions
like 
\begin{equation}  \label{bugs}
<  [e^{ik\phi(x_1)}]_r   \dots   [ e^{ik\phi(x_n)}]_r  >_{\ga}
\end{equation}

We approach the problem by starting again with massive fields and 
then taking the limit as the mass goes to zero  (see also  \cite{Dim98}).
So  as in section \ref{singing}   let   $\{ \phi(f) \} $  with  be a family of Gaussian random variables
indexed  by  $H^{-1}_{\ga, \mu}$   with 
covariance  given by the inner product  $( - \De_{\ga}+ \mu )^{-1} $ and let    $< \cdots >_{\ga, \mu}$
be the expectation.   
  We define first    for smooth $f$  
 \begin{equation}
 [ e^{  i\phi( f)}]_r =   
   e^{  i\phi( f)} e^{   (f, G^\#_{\ga} f) /2}  
 \end{equation}
 where  $ G^\#_{\ga} $ is an operator on  $L^2(\bbC_{\infty}, \mu_{\ga})$ with kernel  $G^\#_{\ga}(x,y)  $
 satisfying
 \begin{equation}
 (f, G^\#_{\ga} h)   =     \int   f(x)  G^\#_{\ga}(x,y)    h(y) d  \mu_{\ga}(x) d  \mu_{\ga}(y) 
 \end{equation} and 
 where  the   kernel    is chosen to have a specific singularity at   $x=y$.  One possible 
 choice is  to take   $G^\#_{\ga}$  to be   $( - \De_{\ga}+ \mu )^{-1} $  in which case the regularization is Wick ordering.    However  the  $\mu$  dependence  leads to problems.  Instead we take something
 which is independent of $\mu$,  has the same singularity, and is still covariant, namely   
 \begin{equation}
 G^\#_{\ga}(x,y)   =  -\frac{1}{2\pi}  \log( d_{\ga}(x,y)  ) 
 \end{equation}
 where  $d_{\ga}(x,y)  $  is the distance.
 
 We want to take   $f= \de_x$ the delta function at  $x$.  Instead let   $\de_{\ka}( \cdot -x)$
 be an approximate delta function in  the plane satisfying  $\int  \de_{\ka}( y -x)dy  =0$.  Then 
  an  approximate  delta function on  $\bbC_{\infty}$  is given  
   (in local coordinates) by  $| \ga|^{-1/2}\de_{\ka}(\cdot -x)$.   Indeed we  have  for any continuous  $h$ 
 \begin{equation}
\lim_{\ka \to \infty}(|\ga|^{-1/2}\de_{\ka}(\cdot -x),h)_{\ga}  = h(x)   \equiv   \de_x(h) 
\end{equation}
We    define  a regularized field
 \begin{equation}
 \phi_{\ka}(z)   =  \phi \left(  | \ga|^{-1/2}\de_{\ka}(\cdot -z)  \right)
 \end{equation}
 and then
  \begin{equation}
 [e^{   ik\phi_{\ka}( z)}   ]_r =   
  e^{   ik\phi_{\ka}( z)}  \exp  \left(    \frac{k^2}{2}  \int   \de_{\ka}(z-x) G^\#_{\ga}(x,y) 
   \de_{\ka}(y-z))  dx dy  \right)
 \end{equation}
 \bigskip
 
\begin{thm}  {  \  }    \label{any}
 Let  
 \begin{equation}
 Z  =  [k_1,z_1,   \dots,  k_n, z_n]
 \end{equation}
 be a sequence of   integers   $k_i$   and points  $z_i  \in \bbC_{\infty}$.
 If the $z_i$  are  distinct  then the limit
\begin{equation}
<Z>_{\ga}   =
\lim_{\ka \to \infty}  \lim_{ \mu   \to    0} 
 <   [e^{   ik_1\phi_{\ka}( z_1)}   ]_r   \cdots  [e^{   ik_n\phi_{\ka}( z_n)}   ]_r  >_{\ga, \mu}  
 \end{equation}
 exists.    
  If  $\ga = e^{\sigma}|dz|^2$
the limit is  
\begin{equation}    \label{goo}
<Z>_{\ga}   
=  \left\{  \begin{array}{lcl}       0 &  &    \sum_i k_i  \neq  0\\
  \exp\left( -\frac{1}{8\pi}   \sum_i k_i^2 \sigma(z_i)  \right)   \prod_{i<j}  |z_i-z_j|^{k_ik_j/2\pi}
  &  &   \sum_i k_i  =  0
  \end{array}   \right.
\end{equation}
\end{thm}
\bigskip

The expression  (\ref{goo})  gives a precise meaning to  (\ref{bugs})  with  $ Z  =  [k_1,z_1,   \dots,  k_n, z_n]$ standing for the formal expression  $ \prod_i [e^{ik_i\phi(z_i)}]_r   $.
Note that        $<Z>_{\ga}   $  is a symmetric function of  the   $(k_i,z_i)$.  The theorem has the immediate Corollary

\begin{cor}  For any conformal metric   $\ga$  
\begin{equation}       \label{googoo}
< Z>_{e^{\sigma}\ga}  =\exp\left(  -  \sum_i \frac{ k_i^2}{8 \pi}  \sigma(z_i)  \right) 
< Z>_{\ga}  
\end{equation}
\end{cor}

Before   proving the theorem we get a preliminary result.      The negative Laplacian is
not invertible on   all of   $L^2(\bbC_{\infty},   \mu_{\ga} )$.  But it is invertible if
we restrict to the orthogonal complement of the constants  denoted  
$L^{2,\perp}(\bbC_{\infty},   \mu_{\ga} )$.  We denote the inverse by  $(-\De_{\ga})^{-1}$.
Also  let
\begin{equation}
G^{\#}(z,z')   =     \frac{-1}{2\pi}   \log|z-z'|   
\end{equation}
be the fundamental solution for  $-\De$  in the plane.

\begin{lem}   {  \  } \label{orange}
\begin{enumerate}
\item   For smooth  $f  \in L^{2,\perp}$
\begin{equation}  
g(z)   =    \int  G^{\#}(z,z')   f(z')  d\mu_{\ga}(z')
\end{equation}
 defines a function on  $\bbC_{\infty}$  which satisfies
$(- \De_{\ga}) g   = f$
\item    For smooth  $f ,h  \in L^{2,\perp}$
\begin{equation}
(h,  (- \De_{\ga})^{-1}  f)   =      \int  h(z)  G^{\#}(z,z')  f(z')      d\mu_{\ga}(z) d\mu_{\ga}(z')
\end{equation}
 \end{enumerate}
\end{lem}
\bigskip

\pr     The  function  $g(z)$ is well-defined  for  $z  \in \bbC$   since the measure  $ d\mu_{\ga}(z') $
is   $\cO(|z'|^{-4})$  as   $z \to \infty$.  We compute
 \begin{equation}
(- \De_{\ga} g)(z)   =  | \ga(z)|^{-1/2} (- \De_z )     \int  G^{\#}(z,z')   f(z')  d\mu_{\ga}(z') 
=    | \ga(z)|^{-1/2}   | \ga(z)|^{1/2}  f(z)    =  f(z)
\end{equation}
 To  include the point at infinity we go to the other coordinate patch.    First    write 
 \begin{equation}
  g(z)   =   \int   ( G^{\#}(z,z')  -   G^{\#}(z,0) )   f(z')  d\mu_{\ga}(z')
  =  \frac{-1}{2\pi}   \int    \log \left|  1-  \frac{z'}{z}  \right|    f(z')  d\mu_{\ga}(z')
\end{equation}
and then   with    $\hat g(\zeta)  =  g(1/\zeta)$   and    $\hat  \mu_{\ga}$
the measure in the new coordinates   we have 
\begin{equation}
 \hat  g(\zeta)       =  \frac{-1}{2\pi}   \int    \log \left|  1-  \frac{\zeta}{\zeta'}  \right|  \hat  f(\zeta')  d \hat \mu_{\ga}(\zeta')
\end{equation}
which is finite at  $\zeta=0$.
Then    $(-  \hat  \De_{\ga})\hat g   =  \hat f$ as before.
This  proves the first point.  

 For the second point   define  $g$ as above and compute
\begin{equation}
((- \De_{\ga})^{-1} h,  f)  =   ((- \De_{\ga})^{-1}  h,(-\De_{\ga} )g) =  (h,g)
\end{equation}
This completes the proof.
\bigskip

\pr  (of  theorem  \ref{any})  
Define  
\begin{equation}
f^{\ga}_{\ka}  =  \sum_i  f^{\ga}_{i, \ka}  \hspace{.5in}          
f^{\ga}_{i \ka}(y)   =    k_i | \ga(y)|^{-1/2}   \de_{\ka}(y - z_i)
\end{equation}
 We have  
 \begin{equation}  \label{sit}
 \begin{split}
 & <   [e^{   ik_1\phi_{\ka}( z_1)}   ]_r   \cdots  [e^{   ik_n\phi_{\ka}( z_n)}   ]_r  >_{\ga, \mu}  
 =    <  \exp(   i\phi( f^{\ga}_{\ka}))>_{\ga,\mu} \exp  \left(    \frac12  \sum_i   (f^{\ga}_{i,\ka},  G^\#_{\ga}  f^{\ga}_{i, \ka}) \right)   \\
    = &  
   \exp  \left(-\frac12  ( f^{\ga}_{\ka}, (-\De_{\ga}+\mu)^{-1}   f^{\ga}_{\ka})
   +  \frac12  \sum_i   (f^{\ga}_{i,\ka},  G^\#_{\ga}  f^{\ga}_{i, \ka}) \right)   \\
 \end{split}
 \end{equation}
Note that     
\begin{equation}
\int    f^{\ga}_{\ka}(y)\ d \mu_{\ga} (y) =   \sum_i  k_i   \int   \de_{\ka}(  y  - z_i) dy  =    \sum_i  k_i 
\end{equation}
If  $\sum_i k_i  \neq 0$   then  $ f^{\ga}_{\ka}$ has  a  constant component  in  $L^2(\bbC_{\infty},  d  \mu_{\ga})$ and hence
\begin{equation}
\lim_{\mu  \to  0 }   ( f^{\ga}_{\ka}, (-\De_{\ga}+\mu)^{-1}  f^{\ga}_{\ka})   =  + \infty
\end{equation}
which shows that the expression (\ref{sit}) goes to zero.  Thus we can restrict attention to 
the case  $\sum_i  k_i  =0$   in which case   $ f^{\ga}_{\ka}$ 
is orthogonal to constants and   $(-\De_{\ga})^{-1}  f^{\ga}_{\ka}$  is well-defined
and  
\begin{equation}
\lim_{\mu  \to    0 }   ( f^{\ga}_{\ka}, (-\De_{\ga}+\mu)^{-1}   f^{\ga}_{\ka})   = 
 (  f^{\ga}_{\ka}, (-\De_{\ga})^{-1}  f^{\ga}_{\ka}) 
\end{equation}
The latter is evaluated by    lemma \ref{orange} and so 
 \begin{equation}
 \begin{split}
  \lim_{\mu \to    0 } <   [e^{   ik_1\phi_{\ka}( z_1)}   ]_r   \cdots  [e^{   ik_n\phi_{\ka}( z_n)}   ]_r  >_{\ga, \mu}   
 =&   \exp\left( - \frac12  \sum_{ij} k_i  k_j \int \de_{\ka}( z_i-x)  G^\#(x,y)  \de_{\ka}(  y  - z_j)   dx dy  
 \right)  \\
    &  
   \exp  \left(  \frac12  \sum_i k_i^2 \int \de_\ka  ( z_i-x  ) 
  G^\#_{\ga}(x,y)  \de_\ka  ( y -z_i  )  dx dy  \right)   \\
 \end{split}
 \end{equation}
Since    $  G^\#(x,y)$ is continuous away from  $x=y$, the terms with  $i \neq j$
have a limit which is   
\begin{equation}
 \exp\left( -   \sum_{i < j} k_i  k_j   G^\#(z_i,z_j)   \right)     =    \prod_{i<j}  |z_i-z_j|^{k_ik_j/2\pi}
\end{equation}
It remains to study the contribution from  terms with  $i=j$  which now have the form
\begin{equation}  \label{seaweed}
 \exp\left( \frac{1}{4\pi}   \sum_i k_i^2\int \de_\ka  ( z_i-x  )\  [  \log|x-y|   -    \log\ d_{\ga}(x,y)   ]\  \de_\ka  ( y -z_i  )  dx dy  \right)  
  \end{equation}
However  $\ga  = e^{\sigma}|dz|^2$ and in  Appendix  \ref{C} we show that
 \begin{equation}
\lim_{y \to x}        [  \log\ d_{\ga}(x,y) -\log|x-y|  ]   = \frac{ \sigma(x)}{2}  
\end{equation}
With this definition at coinciding points   $   [\log|x-y|   -   \log\ d_{\ga}(x,y) ]   $  is 
continuous.    Then    (\ref{seaweed})    has the limit  as  $\ka  \to \infty$
\begin{equation}
 \exp\left( -\frac{1}{8\pi}   \sum_i k_i^2 \sigma(z_i)  \right)  
  \end{equation}
to complete the proof.
\bigskip

Next   we     exhibit the covariance of     our expectations under   Mobius transformations
\begin{equation}
\al(z)   =  \frac { az+b}{cz+d}   \hspace{1in}    ad-bc \neq 0
\end{equation}
\bigskip
These are biholomorphic on  $\bbC_{\infty}$ and   preserve the class of conformal 
 metrics

\begin{lem}   
Let   $\al$  be a Mobius transformation  and $z_i = \al (w_i)$. Then   
\begin{equation}  \label{bing}
<[ k_1,z_1,   \dots,  k_n, z_n]> _{\ga} =        < [k_1,w_1,   \dots,  k_n, w_n]> _{\al^*(\ga)} 
\end{equation}
\end{lem}

\pr    We have  for   the convariance  
\begin{equation}
 (f,  ( - \De_{\ga} +\mu)^{-1} f)_{\ga}
=    (\al^*f,  ( - \De_{\al^*\ga} +\mu)^{-1} \al^*f)_{\al^*\ga}
\end{equation}
and similarly for the operator  $G_{\ga}^\#$.
Thus  we have the identity
\begin{equation}
\begin{split}
&\exp \left(  - \frac12   \sum_{ij}   (f^{\ga}_{i \ka},  ( - \De_{\ga} +\mu)^{-1} f^{\ga}_{j \ka})_{\ga}
+ \frac12    \sum_i    (f^{\ga}_{i \ka},  G^\#_{\ga} f^{\ga}_{i \ka})_{\ga}  \right)\\
=&\exp \left(  - \frac12   \sum_{ij}   (\al^*f^{\ga}_{i \ka},  ( - \De_{\al^*\ga} +\mu)^{-1}\al^* f^{\ga}_{j \ka})_{\al^*\ga}
+ \frac12    \sum_i    (\al^*f^{\ga}_{i \ka},  G^\#_{\al^*\ga}\al^* f^{\ga}_{i \ka})_{\al^*\ga}  \right)\\
\end{split}
\end{equation}
As we have seen  the   left side converges to  $<[ k_1,z_1,   \dots,  k_n, z_n]> _{\ga} $ as   $\mu \to 0, \ka \to \infty$  since      $f^{\ga}_{i \ka}$  converges   to   $k_i \de_{z_i}$ in the metric  $\ga$  when integrated against continuous functions.    For the right
side   first  take   $\mu \to  0$ as before.    Then as      $\ka \to   \infty$  we  have that   
$\al^*f^{\ga}_{i \ka}$  converges to   $k_i  \de_{w_i}$ in the metric  $\al^*\ga$ since for a continuous function     $h$ 
\begin{equation}  \label{sponge}
(\al^*f^{\ga}_{i \ka},h)_{\al^*\ga}    =   
( f^{\ga}_{i \ka}, \al_*h)_{\ga} \  \to   \   k_i  \de_{z_i} (\al_* h)  = k_i  h(w_i)
\end{equation}
Here  $\al_*$ is the push forward  $(\al_*h)(z) =  h( \al^{-1}(z))$. Using this fact one
shows in the same way that the right side converges to  $  <[k_1, w_1,   \dots,  k_n, w_n]> _{\al^*(\ga)} $ to complete the proof.
\bigskip

\res    Given   $Z   = [k_1,z_1, \dots,  k_n, z_n]$  suppose  we take a metric $\ga$   with  $\ga  =  |dz|^2$ near   $z_i$. 
 Then 
\begin{equation}
< Z>_{\ga}  =  \prod_{i<j}  |z_i-z_j|^{k_ik_j/2\pi}
\end{equation}
The  expectation   is     independent   of the metric and we write  it as  
$<Z> $.

If     $\ga  =  |dz|^2$   in some region $\Om$ and    $z = \al(w)$ is a Mobius transformation
then 
\begin{equation}
\al^*( \ga  )   =   \frac{\pa z }{\pa w}   \frac{\pa \bar  z }{\pa \bar   w}  |dw|^2 
 = \left| \frac{\pa z }{\pa w} \right|^2   |dw|^2 
\end{equation}
in  $\al^{-1}(\Om )$.  If we specialize     (\ref{bing})  to an    metric   $\ga$    flat on    $Z$ and use
(\ref{googoo})  on the right side      we get the familiar
\begin{equation}    \label{squash}
<[ k_1,z_1,   \dots,  k_n, z_n]>  =
   \prod_i  \left| \frac{\pa z }{\pa w} (w_i)   \right|^{-k_i^2/4\pi} 
     < [k_1,w_1,   \dots,  k_n, w_n]> 
\end{equation}

\subsection{an algebra of symbols}
We next reformulate some of these results in a more algebraic language.   Define the product of
$Z  = [ k_1,z_1,  \dots,  k_n, z_n ]$ and   $ Z' = [   k'_1,z'_1,  \dots,  k'_m, z'_m]$  to be
  \begin{equation}
Z Z' = [ k_1,z_1,  \dots,  k_n, z_n,  k'_1,z'_1,  \dots,  k'_m, z'_m ]
\end{equation}
Then the space of  sequences   is a monoid and we let 
   $\Upsilon$   be the  free algebra  generated by  this monoid.   The elements of 
$\Upsilon$  are functions  $F$   from  sequences   $Z$  to  $ F(Z)  \in  \bbC$ written
\begin{equation}
F =   \sum_Z F(Z)  Z
\end{equation}
such that   $F(Z) =0$ for all but finitely many  $Z$.
 Scalar multiplication, addition, and multiplication satisfy 
\begin{equation}
\begin{split}
\al F  =&\sum_Z \al F(Z)  Z   \hspace{1cm}    \al  \in \bbC \\
F +  F'   =& \sum_Z  (F(Z)  + F'(Z)) Z\\
FF'  =& \sum_{Z,Z'}  F(Z) F'(Z') ZZ'\\
\end{split}
\end{equation}
We     define  the subspace
\begin{equation}
  \Up_0    =  \{  F \in  \Up:   F(Z)   =0    \textrm{ if  $Z$ has coinciding points }  \}  
  \end{equation}
  and for      $A \subset  \bbC_{\infty}$    the     subalgebra localized in  $A$ 
  \begin{equation}
  \Up_{A}    =  \{  F \in  \Up:   F(Z) =0    \textrm{ if  $Z$ has points outside   $A$ }  \}  
  \end{equation}
Also let
 $\Up_{0,A}=   \Up_0 \cap  \Up_{A} $.

Define  expectations as    a linear functional  on   $ \Up_0   $
by  
\begin{equation}
<F>_{\ga}   =   \sum_Z  F(Z)  <Z>_{\ga}
\end{equation}  
We collect some  properties of these expectations.

\begin{enumerate}

\item   For any  Mobius transformation  $\al$   define   
an isomorphism   $\tau_{\al}$ on  $\Up$ by  setting 
\begin{equation}
 \tau_{\al  }(Z)  =  \tau_{\al } [ k_1,z_1,  \dots,  k_n, z_n] 
=     [ k_1, \al(z_1),  \dots,  k_n, \al(z_n)] 
\end{equation}
and then extending   by linearity to the full algebra. 
Then  for    $F \in \Up_0$
we have by      (\ref{bing})
\begin{equation}  \label{bingo}
<   \tau_{\al} F  >_{\ga}   =    <F>_{\al^*(\ga)}
\end{equation}

\item
Given  two conformal metrics   $\ga',\ga$ with  $\ga' =  e^{\sigma} \ga$  we define
an  isomorphism   $ \tau_{\ga', \ga  }$   on  $\Up$  by 
\begin{equation}
 \tau_{\ga', \ga  }(Z)  =  \tau_{\ga', \ga  } [ k_1,z_1,  \dots,  k_n, z_n] 
=    \exp\left( \frac{1}{8\pi}   \sum_i k_i^2 \sigma(z_i)  \right)      [ k_1,z_1,  \dots,  k_n, z_n] 
\end{equation}
and then extending   by linearity.   Then for    $F   \in \Up_0$    we have by  (\ref{googoo})
\begin{equation}   \label{orca}
<  F> _{\ga} \  = \  < \tau_{\ga'\ga} F > _{\ga'}
\end{equation}

\item   Define a homomorphism $\tau_{ \ka}$   from    $\Upsilon$   to  the polynomial algebra 
generated by  $ [e^{   ik\phi_{\ka}( x)}   ]_r  $   by  
\begin{equation}
\tau_{\ka} ( Z )  =
\tau_{\ka}   (  [ k_1,z_1,  \dots,  k_n, z_n]  )     =     \prod_{i=1}^n  [e^{   ik_i\phi_{\ka}( z_i)}   ]_r  
\end{equation}
and then  extended by linearity.   For  $F \in \Up_0$
we have by theorem     \ref{any}
\begin{equation}
<    F  >_{\ga}   =   
\lim_{\ka \to  \infty}  \lim_{\mu  \to   0} 
<    \tau_{\ka}  ( F ) >_{\ga, \mu}   
\end{equation}

\item
We also   have a reflection positivity result.
Our  radial reflection mapping  $\theta$   induces  a reflection 
$\Theta$  on  sequences by  
  \begin{equation}  \Theta  Z  =
\Theta    [ k_1,z_1,  \dots,  k_n, z_n] =    [ -k_1, \theta z_1,  \dots,  -k_n, \theta z_n]    
\end{equation}
This induces   an  anti-linear mapping   $\Theta$  on   $\Upsilon$ 
 by    
 \begin{equation}
 \Theta  \left(  \sum_Z F(Z)   Z  \right)  =  \sum_Z   \overline{ F(Z)}\   \Theta  Z
 \end{equation}
 We   work  with open discs 
 \begin{equation}
 B_0  =  \{   z:   |z| <1\}      \hspace{1cm}      B'_0 =   \{  z:  |z|  >1\}
 \end{equation}
rather   than the closed discs  $D_0,D'_0$ used in the massive case.   If   $F \in  \Upsilon_{0,B_0}$ then
$\Theta F  \in  \Up_{0,B'_0}$   and   $( \Theta  F)\  F  \in  \Upsilon_0$.
  
 \begin{lem}  
 \label{alone}
  Let   $\ga$ be reflection invariant  $\theta^*\ga = \ga$.     If   $F \in  \Upsilon_{0, B_0}$  
  then 
 \begin{equation}  
 < ( \Theta  F)\  F  >_{\ga}   \  \geq    \  0
 \end{equation}
 \end{lem}
\bigskip

\pr   
First consider  $\Theta $ on functions in  $L^2(Q, \Sigma, m_{\ga, \mu})$
as defined in  (\ref{green}).    Since the power series for  $e^{i \phi(f)}$  converges 
in  $L^2$ and since    $\Theta  \phi(f)  \Theta  =  \phi( \theta^* f)$  we have
$\Theta   e^{i \phi(f)}  \Theta   =   e^{-i \phi(\theta^* f)}$.    Since  $\theta$ is an 
isometry we also have    $(f,  G_{\ga}^\# f) =   (\theta^*f, G_{\ga}^\# \theta^*f)$
and hence    $\Theta   [e^{i \phi(f)}]_r  \Theta   =  [ e^{-i \phi(\theta^* f)}]_r$.

Now we compute  for  $Z,Z'  \in \Up_{B_0}$
\begin{equation}
\begin{split}
<( \Theta Z)Z'>_{\ga} =& <  [-k_1, \theta z_1, \dots,  -k_n,\theta z_n,k'_1,z'_1, \dots,  k'_m, z'_m]>_{\ga}\\
=& \lim_{\ka \to  \infty}  \lim_{\mu  \to   0} 
<  \prod_{i=1}^n  [ e^{-i k_i\phi(\theta^* |\ga|^{-1/2}\de_{\ka} (  \cdot - z_i))}]_r
 \prod_{j=1}^m  [ e^{ik_j \phi( |\ga|^{-1/2}\de_{\ka}  (  \cdot - z_j))}]_r>_{\ga, \mu}  \\
 =& \lim_{\ka \to  \infty}  \lim_{\mu  \to   0} 
< \Theta  \left( \prod_{i=1}^n  [ e^{i k_i\phi(|\ga|^{-1/2}\de_{\ka}  (  \cdot - z_i))}]_r \right)
 \prod_{j=1}^m  [ e^{ik_j \phi( |\ga|^{-1/2}\de_{\ka}  (  \cdot - z_j))}]_r>_{\ga, \mu}  \\
=& \lim_{\ka \to  \infty}  \lim_{\mu  \to   0} 
   < ( \Theta( \tau_{\ka}Z)  \tau _{\ka} Z'>_{\ga, \mu}
\end{split}
\end{equation}
The second step we put     $\theta^* |\ga|^{-1/2}\de_{\ka}  (  \cdot - z_i)$  rather than the usual  $ |\ga|^{-1/2}\de_{\ka}  (  \cdot -  \theta  z_i))$.
This  still works since      $\theta^* |\ga|^{-1/2}\de_{\ka}  (  \cdot - z_i)$
still converges to the delta function at   $\theta z_i$  as in   (\ref{sponge}).

It follows that 
\begin{equation}
 < ( \Theta  F)\  F  >_{\ga}   =   
 \lim_{\ka \to  \infty}  \lim_{\mu  \to   0} 
   <  \Theta ( \tau_{\ka}F)  \tau _{\ka} F>_{\ga, \mu}
\end{equation}
For  $\ka$ sufficiently large $\tau_{\ka} F$ is  $\Sigma_{D_0}$  measurable and so    $  <  \Theta( \tau_{\ka}F)  \tau _{\ka} F>_{\ga, \mu}\geq  0$ by the 
massive result (\ref{sing}).  We conclude  $ < ( \Theta  F)\  F  >_{\ga}  \geq 0$.

\end{enumerate}

\subsection{a standard Hilbert space}

As in section \ref{metrics} we  choose  a standard metric, but now we want it to 
be as  flat is possible.
Well inside  $B_0$   we want it to be     $|dz|^2$.   We  could try to  make it   reflection invariant  by restricting to  $B_0$  and  then  reflecting  to 
get the metric  $|z|^{-4}|dz|^2=  |d \zeta|^2$  in  $B'_0$.  However   the result would not be smooth at the boundary.   This is  corrected   
by    the requirement    that    the metric be     $|z|^{-2}|dz|^2$   on a neighborhood of the boundary.
Thus we define   for some   constant  $d$
  \begin{equation}  \label{standard}
  \ga_0  =  \rho_0(z)  |dz|^2    \ \ \ \ \ \ \ \ \ \ \ \ \ \
\rho_0(z)  =  \left\{ \begin{array}{lcl} 
   1       &  &    |z| <e^{-2d}  \\
    |z|^{-2}    &  &  e^{-d}  <    |z| <e^{d}  \\
     |z|^{-4}     &  &  e^{2d}  <    |z|  
\end{array}
\right.
\end{equation}
In the regions    $  e^{-2d}  \leq |z|  \leq   e^{-d}$  and      $  e^{d}  \leq |z|  \leq   e^{2d}$ 
the function  $\rho_0(z)$ is a smooth interpolation   that  preserves the reflection invariance.

If  $Z,Z'$ are monoids in  $\Up_{0,B_0}$   with  
$Z  = [ k_1,z_1,  \dots,  k_n, z_n ]$   and if   
$d$ is sufficiently small, then by  (\ref{googoo})
 \begin{equation}
       < (    \Theta  Z)\  Z'  > _{\ga_0}  
 =  \prod_i  |\theta z_i|^{k_i^2 /2\pi}       < ( \Theta  Z)\  Z'  > 
 =     < (\tilde    \Theta  Z)\  Z'  > 
\end{equation}
where  we define
   \begin{equation} \tilde   \Theta  Z  =
\prod_i    |\theta z_i|^{k_i^2 /2\pi}     \Theta      Z
\end{equation}
Extend  $\tilde  \Theta$ to be anti-linear on the whole algebra and then 
  for     $F \in   \Up_{0,B_0}$  
 \begin{equation}
 < (\tilde    \Theta  F)\  F  >  =    < (   \Theta  F)\  F  >_{\ga_0}         \  \geq    \  0
 \end{equation}
 by  lemma   \ref{alone}.    This is  positivity  for   the flat expectation. 
      For another way to derive it  see   \cite{FFK89}.
 \bigskip

 Now start with   with the vector space $  \Up_{0, B_0}$ and give  it the norm 
    $\|  F  \| ^2 =  <(\tilde    \Theta F) F>$. Divide by the null  space   $\cN =  \{  F :  \|F\| =0  \}$   and get an
inner product space   $ \cH_{0}    =   \Up_{0,B_0} /\cN     $.
Then  take the completion  to    
get  the standard    Hilbert space    
\begin{equation}
\cH   = \overline{ \cH_{0}}   =   \overline{    \Up_{0, B_0} /\cN }
\end{equation}
If     $\nu$  is  the mapping  from    $\Up_{0,B_0}$  to  $\cH_0$.   
 then  
 \begin{equation}
 (  \nu (  F) , \nu(F'))   =    <(\tilde    \Theta F) F'>
\end{equation}

\subsection{amplitudes}

As in section  \ref{bunko}  we now  suppose we are given open  discs
$B_i   \subset  \bbC_{\infty}$ with disjoint closures  $D_i$.    These will have the form  $B_i = \{ z : |z-a_i| < r_i\}$
or   $B_i = \{ z : |z-a_i| > r_i\}$  or  else  be a half plane.   Let   $\al_i$ be the Mobius 
transformation which takes  $B_i$  to the unit disc  $B_0$ as in   (\ref{butter})  and 
let   $\ga_i  =  \al^*_i \ga_0$   be a standard flat  metric  based on  $B_i$.   We   consider
metrics   $\ga$    such that   $\ga = \ga_i$   on a neighborhood of  $B_i$.   Then   $(\bbC_{\infty}, \ga)$  is
a sphere with standard   flat  discs  $B_i$.

Also as  in    section   \ref{bunko}  we   parametrize   the discs.  They are   divided into  in-discs 
 $\{B_i\}_{i \in I} $ and out-discs  
$\{B_i\}_{i \in I'}$.  For each in-disc  $B_i$   let     $j_i(z) = e^{i \theta_i}\al_i(z)$  which  takes
  $B_i$  to  $B_0$ and   satisfies  and  $j_i^*\ga_0  = \ga_i$.
 For each out-disc  $B_i$  let     $j'_i=e^{i \theta_i}\al_i(z)^{-1}$  which  takes
  $B_i$  to  $B'_0$ and   satisfies  and  $(j'_i)^*\ga_0  = \ga_i$.
These are Mobius transformations and we    define  the isomorphisms on  $\Up$ by 
\begin{equation}
\cJ_i  =   \tau_{j_i^{-1}}   \hspace{1cm}   \cJ'_i  =   \tau_{(j'_i)^{-1}}  
\end{equation}
Then 
\begin{equation}
\cJ_i:    \Up_{B_0}  \to    \Up_{B_i}    \hspace{1cm}   \cJ'_i:    \Up_{B'_0}  \to    \Up_{B_i}    
\end{equation}

Again we want   to study the amplitudes  $Z_{\ga}< \prod_{i  \in I'}  \cJ'_i    \Theta  F_i 
 \prod_{i \in  I} \cJ_iF_i  >_{\ga}$   with   $F_i  \in   \Up_{0,B_0}$
For our purposes it is enough to take   $Z_{\ga} =1$,  although there
there are other possibilities   \cite{Gaw99}.   Also   for  given   $F_i$   we can 
choose  the parameter  $d$ in   $\ga_0$  sufficiently small   so   that  the points of   $F_i$  are entirely in the flat region.
Then  the points  of   $ \cJ_iF_i $ and  $  \cJ'_i    \Theta  F_i $   in  $B_i$   are entirely in the flat region for  $\ga_i$,  and since     
$\ga = \ga_i$ on  $B_i$  the points are in the flat region for   $\ga$.     Then by  (\ref{orca})  the expectations $ < \prod_{i  \in I'}  \cJ'_i    \Theta  F_i 
 \prod_{i \in  I} \cJ_iF_i  >_{\ga}$  
are independent of  the particular   $\ga_0$ or  $\ga$  that we choose.

\begin{thm}   \label{ellen}
 Let   $(\bbC_{\infty}, \ga)$  be  
 a  parametrized sphere with in-discs
  $\{ B_i\}_{i \in I}$   and out- discs   $\{  B_i \}_{i \in I'}$.  
 Then    there is a bilinear form   on the algebraic tensor products  
\begin{equation}
A^{I'I}: \left( \otimes_{i \in I'}  \cH_0 \right)  \times  
  \left(  \otimes_{i \in I} \cH_0   \right)  \to  \bbC 
\end{equation}
 anti-linear in the first factor,    such that  for    $\cF_i =  \nu( F_i),    F_i  \in \Up_{0,  B_0}$
\begin{equation}
  A^{I'I} \left( [\otimes_{i \in I'}  \cF_i],  [\otimes_{i \in I} \cF_i ]\right) =
< \prod_{i  \in I'}  \cJ'_i   \Theta  F_i    \prod_{i\in  I} \cJ_iF_i  >_{\ga}
\end{equation}\end{thm}
\bigskip

\pr  
For   any   $k\in  I$  let    $G_k  =  \prod_{i  \in I'}  \cJ'_i    \Theta  F_i 
 \prod_{i \in  I, i\neq  k } \cJ_iF_i  $.
Then we have  
\begin{equation} 
\begin{split}
&< \prod_{i  \in I'}  \cJ'_i    \Theta  F_i    \prod_{i\in  I} \cJ_iF_i  >_{\ga}
=  <G_k \cJ_kF_k  >_{\ga}\\
= & <\tau_{\ga_k,\ga}(G_k \cJ_kF_k ) >_{\ga_k}
= <(\tau_{\ga_k,\ga}G_k)   \cJ_kF_k ) >_{\ga_k}
=  <(\cJ_k^{-1}\tau_{\ga_k,\ga}G_k)  F_k  >_{\ga_0}\\
\end{split}
\end{equation}
Here we have  used  (\ref{orca}),  then   $\tau_{\ga_k,\ga}( \cJ_kF_k )  =  \cJ_k F_k$
since   $\ga = \ga_k$ on  $B_k$,  then  (\ref{bingo}).
Note that   $\tau_{\ga_k,\ga}G_k  \in  \Up_{0,B'_k}$
so    $\cJ_k^{-1}\tau_{\ga_k,\ga}G_k  \in  \Up_{0,B'_0}$.
 Then     by  the Schwarz inequality for the bilinear form   $< (\Theta F)F'>_{\ga_0}$  we have
\begin{equation}  \label{stung}
|< \prod_{i  \in I'}  \cJ'_i    \Theta  F_i    \prod_{i\in  I} \cJ_iF_i  >_{\ga}|   
\leq \|  F_k\|  \|    \Theta \cJ_k^{-1}\tau_{\ga_k,\ga}G_k \|
\end{equation}
where $\|F \|^2  =   < (\Theta F)F>_{\ga_0}  =  < (\tilde  \Theta F)F>  $.  
Hence the expression   only depends on the equivalence class  of  each  $F_k$  and we have
a linear functional on   $\cH_0$.  The argument
is   similar   for   $k \in  I'$ but we get an  anti-linear functional on  $\cH_0$.

Now   we have a multilinear functional  on   $ \left( \times_{i \in I'}  \cH_0 \right)  \times 
  \left(  \times_{i \in I} \cH_0   \right)$, anti-linear  in the first group of  factors,    and
this gives a  mapping  $ \left( \otimes_{i \in I'}  \cH_0 \right)  \otimes 
  \left(  \otimes_{i \in I} \cH_0   \right)  \to  \bbC$  by the universal property of the tensor product  \cite{Gre78}.  Hence it also determines a bilinear form  on  $ \left( \otimes_{i \in I'}  \cH_0 \right)  \times  
  \left(  \otimes_{i \in I} \cH_0   \right)$.
This completes the proof.    
\bigskip

\res  This result is considerably weaker than the massive result.  The basic  space
is  the pre-Hilbert space  $\cH_0$  not the  completion  $\cH$.  The tensor
product is the algebraic tensor product not the Hilbert space tensor product.   And 
$\cA^{I'I}$  is a bilinear form rather than an operator.

Note  that   
as a consequence  of  (\ref{stung})  the functional    is continuous in 
any  particular variable   $F_k$  and one can extend  the definition from  $\cH_{0}$  to the
completion $\cH$.    But  it is not proved that one can do it for all   $F_k$  at   once
  and  so   we    do not have a bounded linear  functional on
 $\cH \times  \dots  \times  \cH$,  let alone a Hilbert-Schmidt functional.  These are  the obstacles  to obtaining the stronger
 results of the massive case.

       Since the algebraic tensor product    is not complete the bilinear form does not necessarily 
define an operator.    However   it does define an operator if  there is only one in-disc  or only one out-disc.     For example   $A^{1I}(  \cF,   \otimes_{i \in I}\cF_i)$
is defined and continuous in  $\cF  \in   \cH$  as noted above,  and  so by the Riesz theorem  there
is  a  linear operator    $A^{1I}:  \otimes_{i \in I}\cH_i  \to  \cH$ such that   
\begin{equation}
  (\cF,   A^{1I}  [\otimes_{i \in I}\cF_i])=A^{1I}(  \cF, [  \otimes_{i \in I}\cF_i]) 
  \end{equation}
Similarly there  is    a  linear   operator     $[A^{I'1}]^*:   \otimes_{i \in I'}\cH_i  \to  \cH   $
such that   
\begin{equation}
  ( [ A^{I'1} ]^* [\otimes_{i \in I'}\cF_i],  \cF  )=A^{I'1}(  \otimes_{i \in I'}\cF_i,\cF) 
  \end{equation}
But   $ A^{I'1} $  itself is not defined.

\subsection{sewing}

We   study  the   amplitudes   $ A^{1I}$ with one out disc.   For  simplicity we   assume the in-discs  $\{B_i\}_{i \in I}$ are all in  $B_0$ and the out disc is  $B'_0$  with the identity parametrization.   Then  our metric 
will satisfy    $\ga  =  \ga_0$    on a neighborhood of   $B'_0$.  
  Take    $ \cF   =  \nu(F)$  and    $\cF_i =  \nu(F_i)$   and   compute   
\begin{equation}  \label{dec}
\begin{split}
\left(  \cF,   A^{1I}  [\otimes_{i \in I} \cF_i ]\right) 
=&
<   \Theta  F    \prod_{i  \in  I}\cJ_i F_i  >_{\ga}  \\
 =&   <   \Theta  F \  \prod_{i \in I}   \tau_{\ga_0,\ga}\cJ_iF_i     > _{\ga_0}  \\
=&(  \cF,   \nu    ( \prod_{i \in I}\tau_{\ga_0,\ga}\cJ_iF_i   )) 
\end{split}  
\end{equation}
where we use     $\tau_{\ga_0,\ga}  \Theta F=   \Theta F$.
Since the   $\cF =  \nu(F)$ are dense we conclude 
   \begin{equation}  \label{eve}
A^{1I}   
 [\otimes_{i \in I} \cF_i ]   = \nu    \left( \prod_{i \in I}\tau_{\ga_0,\ga}\cJ_iF_i  \right)
 \end{equation}

We also  study amplitudes $A^{I'1}$ with one in-disc.
We assume the out-discs are all in    $B'_0$ and the in-disc is  $B_0$ with identity parametrization.   
Then  
 $\ga  =  \ga_0$  on a neighborhood of  $B_0$  
    Take    $ \cF   =  \nu(F)$  and    $\cF_i =  \nu(F_i)$.   
and compute    
 \begin{equation}
\begin{split}
\left([A^{I'1}  ]^*  [\otimes_{i \in I'} \cF_i ],\  \cF \right) 
=&
<   ( \prod_{i \in I'}   \cJ'_i \Theta F_i  )\  F   >_{\ga}  \\
=&
<   ( \prod_{i \in I'}    \tau_{\ga,\ga_0} \cJ'_i    \Theta F_i  )\  F   >_{\ga_0}  \\
=&(  \nu  ( \Theta  \prod_{i \in I'}   \tau_{\ga_0,\ga} \cJ'_i   \Theta F_i   ) , \cF)   \\
\end{split}
\end{equation}
and so    
   \begin{equation}  \label{march}
(A^{I'1} )^*   
 [\otimes_{i \in I'} \cF_i ]   =  \nu  \left( \Theta  \prod_{i \in I'}   \tau_{\ga_0,\ga} \cJ'_i    \Theta F_i  \right) 
  \end{equation}

Now we state the sewing result with a simple configuration.  (Compare theorem  \ref{over}).

\begin{thm}
Let   $A^{1I}$  be the amplitude for  transitions from   $\{ B_i \}_{i \in I}$  in   $B_0$   to  $B'_0$, the latter
with identity parametrization. 
 Let   $A^{I'1}$  be the amplitude for  transitions from  $B_0$ with identity parametrization   to 
  $\{ B_i \}_{i \in I'}$ in  $B'_0$.   Finally let     $A^{I'I}$
  be the amplitude for  transitions from   $\{ B_i \}_{i \in I}$  to 
  $\{ B_i \}_{i \in I'}$  with the same parametrizations.
Then  
 $A^{I'1} A^{1I}   =   A^{I'I}$ in the sense that 
\begin{equation}
  \left([ A^{I'1} ]^*  [\otimes_{i \in I'} \cF_i ],    A^{1I} [\otimes_{i \in I} \cF_i ]\right)
  =A^{I'I} \left(  [\otimes_{i \in I'} \cF_i ],   [\otimes_{i \in I} \cF_i ]\right) 
   \end{equation}
 \end{thm}

\bigskip

 \pr   Let   $\ga_1$ be a metric suitable for   $A^{1I}$  so that   $\ga_1 = \ga_0$  on
 a neighborhood of  $B'_0$  and   $\ga_1 =  \ga_i$  on  $B_i,  i \in I$.
 Let   $\ga_2$   be a metric suitable for   $A^{I'1}$  so that   $\ga_2 = \ga_0$  on
 a neighborhood of  $B_0$  and   $\ga_2 =  \ga_i$  on  $B_i,  i \in I'$.
 Define a smooth metric  $\ga$ by   $\ga =  \ga_1$ in   $B_0$  and   $\ga= \ga_2$
 in   $B'_0$.   Then   $\ga$ is a suitable metric for   $A^{I'I}$ and we compute
   using   (\ref{eve}), (\ref{march})
\begin{equation}    \label{feb}
\begin{split}
 \left( [ A^{I'1} ]^*  [\otimes_{i \in I'} \cF_i ],   A^{1I} [\otimes_{i \in I} \cF_i ]\right)
=&  \left(   \nu  (  \Theta  \prod_{i \in I'}  	\tau_{\ga_0,\ga_2} \cJ'_i   \Theta F_i  ) ,
  \nu   ( \prod_{i \in I}	\tau_{\ga_0,\ga_1}\cJ_iF_i   )  \right)   \\
   =&  <    \prod_{i \in I'}   	\tau_{\ga_0,\ga_2} \cJ'_i    \Theta F_i  
     \prod_{i \in I}	\tau_{\ga_0,\ga_1}\cJ_iF_i      >_{\ga_0}   \\
   =&  <      \prod_{i \in I'}    	\tau_{\ga_0,\ga} \cJ'_i    \Theta F_i \ 
       \prod_{i \in I}	\tau_{\ga_0,\ga}\cJ_iF_i    >_{\ga_0}   \\
       =&  <    \prod_{i \in I'}    \cJ'_i    \Theta F_i     \prod_{i \in I}(\cJ_iF_i)       >_{\ga}   \\
     =&A^{I'I} \left(  [\otimes_{i \in I'} \cF_i ],   [\otimes_{i \in I} \cF_i ]\right) \\
\end{split}
\end{equation}
 \bigskip

 \res
 \begin{enumerate}
 \item   Much more general configurations are possible using the same basic ideas.
 \item
   Somewhat similar results  have been obtained by Tsukada
\cite{Tsu91},      however in this work the formulation of the problem is 
rather different.
\end{enumerate}
 
     \newpage

\appendix
\section{Wick monomials}  \label{A}
 We  state some results about Wick monomials on the sphere. 
 \footnote{All the results in the appendices  hold with the sphere replaced by  a compact two dimensional    manifold.} 
These  are   generalizations of standard results in the plane  \cite{Si75},  \cite{GJ87}.  However we avoid regularizing the field with approximate delta functions.

Let  $m_C  =m_{\ga, \mu}$ 
 be the Gaussian measure  with covariance  $ C =      ( - \De_{\ga} + \mu )^{-1}$   as in the 
 text.
 We  want to  consider expressions of the form  
\begin{equation}  \label{stun}
V=  \frac12  \int   :\phi(x)   \phi(y):  v(x,y) d\mu_{\ga}(x)   d \mu_{\ga} (y)
\end{equation} 
where formally     $\phi(x)  =  \phi(\de_x)$. We take     $v(x,y)$  to be   the distribution  kernel of a bounded    symmetric  bilinear form     operator   $v$ on  $H^{+1} \times H^{+1}$, i.e.  
\begin{equation}
v(f,f')   =   \int   v(x,y)  f(x) f'(x)  d\mu_{\ga}(x)   d \mu_{\ga} (y)
\end{equation}
If   $e_i$ is an orthonormal basis for  $H^{+1}$  and   $\chi^i = C^{-1}e_i$  is the dual basis 
for   $H^{-1}$   then   $f=   \sum_i  e_i (\chi^i , f)  $  and so we 
have   
\begin{equation}
v(f,f')   =   \sum_{i,j}  v(e_i,e_j) (\chi^i ,f) ( \chi^j ,f')  
\end{equation}
Thus  what we seek to define is
\begin{equation}
V= \frac12  \sum_{i,j}   :\phi(\chi^i )   \phi(\chi^j ): v(e_i,e_j)
\end{equation}
As an approximation we consider   
 \begin{equation}
V_N  =  \frac12   \sum_{1 \leq  i,j \leq N}    :\phi(\chi^i )   \phi(\chi^i ): v(e_i,e_j)
   \end{equation}
which is well-defined.

\begin{lem}   If    $\sum_{ij} |v(e_i,e_j)|^2$  converges then   
\begin{enumerate}
\item
$V=  \lim_{N \to \infty}  V_N$
exists  in      $L^2(Q, \Sigma,  m_C)$  and satisfies 
  \begin{equation}  \label{skunk}
 \|V\|^2_2  = \frac12  \sum_{ij} |v(e_i,e_j)|^2
 \end{equation}
 \item   For  $h_i \in  H^{-1}$
 \begin{equation}  \label{tree}
 \int  :\phi(h_1) \cdots \phi(h_n):  V  d m_C  =
 \left\{   \begin{array}{rcl}
 0 &   &   n \neq 2  \\
 v(Ch_1,Ch_2)  &   &    n=2  
 \end{array} 
 \right.
 \end{equation}
 \item  The definition of $V$ is independent of   basis.
 \end{enumerate}
\end{lem}
\bigskip

\re  Note that 
the bilinear form  $v$   on  $H^{+1} \times  H^{+1}$determines a bounded operator $v$    from  $ H^{+1} $ to $H^{-1}$  so that  
\begin{equation}
v(f,g)   =   (f,  vg)_{+1,-1}
\end{equation}
Using also     
\begin{equation}
(f,h)_{+1,-1}  =  (C^{-1}f,h)_{-1}  =(f,Ch)_{+1}
\end{equation}
 we have  
 \begin{equation}
\sum_{ij} |v(e_i,e_j)|^2=  \sum_{ij} |(e_i,ve_j)_{+1,-1}|^2  =  \sum_{ij}|(\chi^i , v e_j)_{-1}|^2
=    \sum_{i}\| v e_j\|_{-1}^2
\end{equation}
Thus the condition  $\sum_{ij} |v(e_i,e_j)|^2 <  \infty$  is the same as the condition that 
the operator $v$ be Hilbert-Schmidt and we have     
 \begin{equation}
 \|V\|^2_2  = \frac12  \|v\|^2_{HS}
 \end{equation}
 An  equivalent  condition is   that
 $C^{\frac12}vC^{\frac12}$ is Hilbert-Schmidt on  $H^0$    ($C^{1/2}$ is unitary from   $H^0$ to  $H^{+1}$
 and from $H^{-1}$ to  $H^0$).  Another    equivalent  condition is   that
 $vC$ is Hilbert-Schmidt on  $H^{-1}$    ($C$ is unitary from   $H^{-1}$ to  $H^{+1}$).
 \bigskip

\pr   We compute  for  $M >N$
\begin{equation}
\begin{split}
\| V_M - V_N  \|^2   =&
\frac14    \int   |  \sum_{N \leq i,j \leq  M}  :\phi(\chi^i )   \phi(\chi^j ): v(e_i,e_j)|^2  d m_C\\
 =&
 \frac12    \sum_{N \leq i,j \leq  M}  | v(e_i,e_j)|^2  
   \\
\end{split}
\end{equation}
This converges to zero  as  $N,M \to \infty$  so  $V$ exists.  The identity  
(\ref{skunk}) is established similarly.

For the second point  we note that   $V$ is in the  closed   subspace spanned   by 
the quadratic Wick monomials
$:\phi(h)  \phi(h'):$.  Thus it is orthogonal to Wick monomials of any other degree.
We compute  
\begin{equation}
\begin{split}
 \int  :\phi(h)  \phi(h'):  V\  d m_C 
 = & \lim_{N \to \infty}  \int  :\phi(h)  \phi(h'):  V_N\  d m_C   \\
=  &  \lim_{N \to \infty}   \sum_{1 \leq  i,j \leq N}   (\chi^i ,h)_{-1}   (\chi^j ,h')_{-1}  v(e_i,e_j)\\
=  & \lim_{N \to \infty}    \sum_{1 \leq  i,j \leq N}   (\chi^i ,Ch)_{-1,+1}   (\chi^i ,Ch')_{-1,+1}  v(e_i,e_j)\\
=  &      v(Ch,Ch')\\
\end{split}
\end{equation}

For the third point note that   the inner products of  $V$  with Wick monomials
are independent of basis by  the previous result.   Since  the Wick monomials span
a dense set it  follows that  $V$ is independent of basis.  This completes the proof.
\bigskip

   In the text we are particularly concerned with   the case  where
the bilinear form  has 
 the kernel      (in local coordinates)    
\begin{equation}   v(x,y)  =  | \ga(x)|^{-1/2}g(x)  \de(x-y)
\end{equation}
for some smooth function  $g$ on  $\bbC_{\infty}$.  Thus the bilinear form is 
\begin{equation}  
v(f,f')   =    \int   f(x)  f'(x)  g(x)   d \mu_{\ga} (x)
\end{equation}
The associated operator  from   $H^{+1}$    to  $H^{-1}$ is just multiplication by  $g$   
and it is Hilbert-Schmidt since   is is bounded on $H^{+1}$  and the injection
  $H^{+1} \to  H^{-1}$  is  Hilbert-Schmidt.  (The latter   is equivalent to the statement 
that   $C$ is Hilbert-Schmidt on  $H^0$.)  Thus $V$ exists in this case and we denote
it by  
\begin{equation}
\frac12 :\phi^2:(g)   =   \frac12 \int   : \phi(x)^2:  g(x)  d \mu_{\ga} (x)
\end{equation}

\begin{lem}  \label{omega}
If  $A$ is closed in  $\bbC_{\infty}$ and   
   $\supp \  g  \subset   A$  then   
$:\phi^2:(g)$   is $\Sigma_A$  measurable.
\end{lem}   
\bigskip

\pr   It suffices to show  $\cE_A ( :\phi^2:(g))=:\phi^2:(g)$.   This follows if the
inner product with any Wick monomial is the same,  and it suffices to consider quadratic monomials.
    Taking the conditional 
expectation onto the Wick monomial we must show 
\begin{equation}  \label{ollie}
   \int  :\phi(e_Ah)  \phi(e_Ah'):  \  :\phi^2:(g) d m_C
= \int  :\phi(h)  \phi(h'):  \  :\phi^2:(g) d m_C 
\end{equation}
But  by  (\ref{tree}) and      $e_A g=g$     the left side of  (\ref{ollie}) is computed as    
\begin{equation}
( Ce_Ah,  g  Ce_Ah)_{+1,-1}
=   (e_Ah,  g  Ce_Ah)_{-1}
= (h, g  Ce_Ah)_{-1}=
( Ch,  g  Ce_Ah)_{+1,-1}
\end{equation}
Similarly the other  $e_A$ is eliminated and we get  $(Ch,  g  Ch)_{+1,-1}$
which is the evaluation of the right side of  (\ref{ollie}).  This completes the proof.

\section{perturbations of Gaussian measures}
  \label{B}

In this section we  consider  quadratic    perturbations of Gaussian measures.
The   treatment is a slightly different  formulation of standard results  \cite{Si75},  \cite{GJ87}.

 Again we consider  the Gaussian measure  $m_C  =m_{\ga, \mu}$  with covariance  $ C =      ( - \De_{\ga} + \mu )^{-1}$   as in the 
 text.  We   want to study    measures of  the    form  $e^{-V}  d m_C$  where       $V$ is a Wick monomial
defined by a bilinear form   $v$  on   $H^{+1} \times   H^{+1}$ as in   (\ref{stun}).
  We continue to  assume that  $v$ satisfies a   Hilbert-Schmidt  condition   (i.e.  $vC$ on $H^{-1}$
  is Hilbert-Schmidt) so that   $V$ exists. We also need     a positivity condition  which is
 \begin{equation}  \label{pos}
  \|f\|^2_{+1}  + v(f,f) > 0  \hspace{1cm}    f \in  H^{+1}
\end{equation}   
 or equivalently
 \begin{equation}
 -\De_{\ga}  + \mu  + v   >0
 \end{equation}
  Note that  the operator $vC$ on $H^{-1}$  is self-adjoint since  
 \begin{equation}  \label{gaga}
(h', [vC] h)_{-1}  =   (Ch', v Ch)_{+1, -1}  =   v(Ch', Ch)
\end{equation}
  Furthermore  
the positivity condition implies  
\begin{equation}
(h, [vC] h)_{-1} = v(Ch, Ch) > - \|Ch\|^2_{+1}  =   - \|h\|^2_{-1} 
\end{equation}
so we have  $vC > -1$.

 \begin{lem}    Let  $ vC$    
be  Hilbert-Schmidt   and  suppose    the positivity condition (\ref{pos}) is satisfied .
 \begin{enumerate}
 \item
  $  e^{-V}$    is  integrable with 
respect to   $m_C$ 
and  
\begin{equation}
\int  e^{-V}  d  m_C  =  { \det }_2 ( 1+ vC)
\end{equation}
\item  
For   $h \in  H^{-1}$     
\begin{equation}
\frac{   \int     e^{i \phi(h)}e^{-V}  d  m_C}{\int  e^{-V}  d  m_C}
=  \exp  \left( - \frac12   (h, (1+vC)^{-1}h)_{-1}  \right)
\end{equation}
Hence  the    measure  $e^{-V}  d  m_C$,  once normalized,  is Gaussian with covariance  $(1+vC)^{-1}$
\end{enumerate}
\end{lem}
\bigskip

\re
Note that   the   variance    of the measure   can also be characterized
as   
\begin{equation}
 (h, (1+vC)^{-1}h)_{-1}   =    (h, C(1+vC)^{-1}h)_{-1,+1}   
=     (h,(C^{-1}+v)^{-1}h)_{-1,+1}    
\end{equation} 
which is the same as   
\begin{equation}
  (h,  (\De_{\ga} + \mu + v)^{-1} h)   
\end{equation} 

\bigskip

\pr  Pick an   orthonormal basis
$\chi^i$  of  $H^{-1}$ consisting of   eigenvectors   for  $vC$ so  $[vC]\chi^i =  \la_i  \chi^i$ with  $\la_i > -1$
and   $\sum_i \la_i^2 < \infty$.
Let    $e_i= C\chi^i$ be the dual  orthonormal basis  for   $H^{+1}$.
Then   we have   from (\ref{gaga})
\begin{equation}
 v(e_i,e_j) = (\chi_i,[vC] \chi_j)_{-1} =  \la_i \de_{ij}
 \end{equation}
  We compute 
\begin{equation}
\begin{split}
\int  e^{-V_N}  d  m_C
=&    \int   \exp  \left(   - \frac12     \sum_{i=1}^N \la_i  :\phi(\chi_i) ^2 :  \right)
\  dm_C  \\
=&      \prod_{i=1}^N \frac{  \int   \exp  (   - \frac12   \la_i ( x_i^2- 1 ) - \frac12 x_i^2   )  \ dx_i }
{  \int   \exp  (   - \frac12 x_i^2   )  \ dx_i }
\\
=&      \prod_{i=1}^N   (1 + \la_i)^{-1/2} e^{\la_i/2} \\
\end{split}
\end{equation}
 Since  $ \log( (1+ \la_i)^{-1/2}e^{\la_i/2})  =  \cO(\la_i^2)$
 and since $\sum_i  \la_i^2 < \infty$   this 
 has a limit as  $N \to \infty$ which is $  {\det}_2  (I   +  vC)^{-1/2}$  where 
 \begin{equation}
  {\det}_2  (I   +  vC)  \equiv   \prod_{i=1}^{\infty}  (1+ \la_i)e^{-\la_i}
  \end{equation}
   Thus we
have  
\begin{equation}  \label{smelt}
\lim_{N \to \infty}  \int  e^{-V_N}  d  m_C
=   {\det}_2  (I   +  vC)^{-1/2}
\end{equation}

Next we note that  for  $M>N$   \begin{equation}
\begin{split}
\|e^{-V_M/2}  - e^{-V_N/2}  \|_2^2
=&  \int  (e^{-V_M}  +e^{-V_N}  - 2 e^{-(V_M +V_N)/2}  ) d  m_C    \\
=&
    \prod_{i=1}^M   (1 + \la_i)^{-1/2} e^{\la_i/2}   +     \prod_{i=1}^N   (1 + \la_i)^{-1/2} e^{\la_i/2} \\
-  &  2    \prod_{i=1}^N   (1 + \la_i)^{-1/2} e^{\la_i/2}     \prod_{i=N+1}^M   (1 + \la_i/2)^{-1/2} e^{\la_i/4} \\
\end{split}
\end{equation}
Hence   it converges to zero   as  $M,N \to \infty$.
Then   
\begin{equation}
   \|e^{-V_M}  - e^{-V_N}  \|_1  \leq   \|e^{-V_M/2}  - e^{-V_N/2}  \|_2\|e^{-V_M/2}  + e^{-V_N/2}  \|_2
\end{equation}
goes to zero as well  and hence     $e^{-V_N}  $  converges in  $L^1$.    Since
a subsequence of   $V_N$  converges pointwise almost everywhere to  $V$, we  conclude
that  $e^{-V}   \in L^1$  and that      $e^{-V_N}  \to  e^{-V}$     in  $L^1$.  Combined with  (\ref{smelt})
this establishes   that  $ \int  e^{-V}  d  m_C
=   {\det}_2  (I   +  vC)^{-1/2}$.  This completes the first 
part

For the second part  expand  for  $h \in H^{-1}$
in   the orthonormal basis  $\chi^i$ we 
by    $h  =  \sum_{i}h_i \chi^i$ with   $\sum_i h_i^2 < \infty$.   Also  define  the approximation  $h_N  =  \sum_{i=1}^Nh_i \chi^i$ .      
Then we have     
\begin{equation}
\begin{split} 
\frac{\int  e^{i\phi(h_N)} e^{-V_N}  d  m_C}{\int  e^{-V_N}  d  m_C} 
=&    \frac{    \prod_{i=1}^N \int        \exp  \left( ih_ix_i  - \frac12  \la_i( x_i^2 - 1 ) - \frac12 x_i^2 \right)  dx_i}
{  \prod_{i=1}^N \int        \exp  \left(  - \frac12  \la_i( x_i^2 - 1 ) - \frac12  x_i^2 \right)  dx_i   }  \\
=&   \prod_{i=1}^N       \exp  \left(  - \frac12  h_i^2  (1 + \la_i)^{-1} \right)            \\
\end{split}
\end{equation}
Taking the limit $N \to \infty$ 
this converges to 
\begin{equation}
\frac{\int  e^{i\phi(h)} e^{-V}  d  m_C}{\int  e^{-V}  d  m_C} 
=       \exp  \left(    - \frac12   \sum_i     h_i^2  (1 + \la_i)^{-1} \right)   =
     \exp  \left( - \frac12   (h, (1+vC)^{-1}h)_{-1}  \right)
\end{equation}
as announced.

\section{distances in conformally equivalent metrics}  \label{C}

Let   $\ga, \ga'$  be conformal metrics on the  Riemann sphere.
The  following    lemma compares distances as points come together.

 \begin{lem}   \label{apple}
  If  $\ga'  = e^{\sigma} \ga$ then
\begin{equation}
\lim_{y \to x}        \frac{ d_{\ga'}(x,y)} {d_{\ga}(x,y)}   = e^{ \sigma(x)/2}  
\end{equation}
\end{lem}
\bigskip

\pr     For each  $y$  near $x$  let  $\al_{xy}(t)$  be the geodesic for  $\ga'$ with 
$\al_{xy}(0) =x$ and  $\al_{xy}(1)  =y$.   In terms of the exponential map  at $x$ it
has the representation
 \begin{equation}  \label{olive}
 \al_{xy} (t) =  \exp_x ( t \exp_x^{-1} (y) )
 \end{equation}
If  $\ga  =  \rho |dz|^2$  then     $\ga'  = e^{\sigma} \rho |dz|^2$  and   we have   
 \begin{equation}
 d_{\ga'} (x,y) =  \int _0^1    e^{\sigma (\al_{xy}(t)) /2 }\sqrt{\rho(\al_{xy}(t) )}  |\al'_{xy}(t) |  dt
\end{equation}
 Given  $\ep >0$   choose $\de_0$    so that   if  $|x-y|  <  \de_0$  then  
 $|\sigma(x) - \sigma(y)|  <  \ep$.  Then choose  $\de_1$  so that if   $|x-y| < \de_1$
 then   $|\al_{xy}(t)  -  x|  <  \de_0$  for all  $0 \leq   t \leq 1$.  Since   $x =  \al_{xx}(t)$
 this is continuity  of  $\al_{xy}(t)$  as  $y \to x$   uniformly in $t$.   That  one can
 accomplish  this follows from the representation   (\ref{olive})
and  the
continuity of the exponential map and its inverse.
 Now  for   $|x-y| < \de_1$
\begin{equation}
\begin{split}
d_{\ga'} (x,y) = &e^{\sigma(x)/2} \int _0^1    e^{(\sigma (\al_{xy}(t)) - \sigma(x) )/2 }\sqrt{\rho(\al_{xy}(t) )}  
 |\al'_{xy}(t) |  dt\\
\geq   & e^{\sigma(x)/2} e^{-\ep/2} \int _0^1  \sqrt{\rho(\al_{xy}(t) )}   |\al'_{xy}(t) |  dt\\
\geq   & e^{\sigma(x)/2} e^{-\ep/2}\ d_{\ga} (x,y) \\
 \end{split}
\end{equation}
Reversing the roles   of the metrics   there is a   $\de_2$ such that if  $|x-y| < \de_2$ 
\begin{equation}
d_{\ga}(x,y)     \geq   e^{-\sigma(x)/2} e^{-\ep/2}\ d_{\ga'} (x,y) 
\end{equation}
Hence  if  $\de  =  \min( \de_1, \de_2 )$  and  $|x-y| <\de$ then    
\begin{equation}
  e^{-\ep/2}    \leq    \frac   { d_{\ga'} (x,y)}  { e^{\sigma(x)/2}   d_{\ga} (x,y)}  \leq 
e^{\ep/2}
\end{equation}
which is the result.


\bigskip


\newpage

\begin{thebibliography}{99}
\bibitem{DDD86} G. De Angelis, D.   de Falco,  G. Di Genova,     
Random fields on Riemannian manifolds: a constructive approach.
Commun. Math. Phys. \textbf{103}, 297-303, 1986



\bibitem{Dim98}  J. Dimock,  Bosonization of massive fermions,  Commun. Math. Phys.  \textbf{198}, 247-281,
1998.


\bibitem{Dim04}  J. Dimock,  Markov quantum fields on a manifold, Rev. Math. Phys \textbf{16},
243-255, 2004






\bibitem{FFK89}  G. Felder, J. Frohlich, J. Keller, On the structure
of unitary conformal field theory, Commun. Math. Phys. \textbf{124}, 417-463, 1989




\bibitem{Gaw99}  K. Gawedski, Lectures on conformal field theory,  in {\em Quantum fields and strings: a course for 
mathematicians},   P. Deligne et. al. eds., American Mathematical Society, Providence, 1999. 


\bibitem{GJ87}  J. Glimm, A. Jaffe, {\em  Quantum physics, an functional integral
point of view}. Springer, 1987.


\bibitem{Gre78}  W. Greub, {\em Multilinear Algebra},  Springer,  1978.


\bibitem{Hua97}  Y.Z. Huang,  {\em  Two  dimensional conformal geometry and vertex 
operator algebras},  Birkhauser, Boston, 1997. 

\bibitem{JaRi06}  A. Jaffe,  G. Ritter,    Quantum field theory on curved backgrounds,  hep-th/0609003.


\bibitem{KaRi83}  R. Kadison, J. Ringrose,  {\em Fundamentals of the theory of operator algebras, Vol 1},
Academic Press, New York, 1983.



\bibitem{Nel73b} E. Nelson,  The free Markov field,  J. Func. Anal. \textbf{12}, 211-227, 1973.


\bibitem{Nel73c}  E. Nelson, Probability theory and Euclidean field 
theory,
in G. Velo., A. Wightman, eds, {\em  Constructive Quantum field theory}, Springer, 1973.


\bibitem{OS73}   K. Osterwalder,  R. Schrader,   Axioms for Euclidean Green's functions,
Commun. Math. Phys. \textbf{31}  83-112, 1973  and   Commun. Math. Phys. \textbf{42}  281-305, 1975. 


\bibitem{Seg89}  G. Segal,  Two-dimensional conformal field theories and modular 
functions, {\em IX International Congress on Mathematical Physics}, B. Simon, A. Truman,
and I.M. Davies eds., 22-37, Adam Hilger, 1989.


\bibitem{Seg04} G. Segal,   The definition of conformal field theory,  in   { \em Topology, geometry, and
quantum field theory},  U. Tillman, ed., Cambridge University  Press,  2004.



\bibitem{Shu01}  M. Shubin, {\em Pseudodifferential operators and spectral theory},    Springer,
2001.


\bibitem{Si75}  B. Simon, {\em  The $P(\phi)_2$ Euclidean field theory}, Princeton University Press,
Princeton, 1974.



\bibitem{Tsu91}  H. Tsukada,   {\em String path integral realization of vertex operator algebras},
Memoirs of the AMS,  Providence,  \textbf{91}, 1991. 



\end{thebibliography}
\end{document}